\documentclass[a4paper,twocolumn,10pt,accepted=2026-01-07]{quantumarticle}
\pdfoutput=1
\usepackage{amsmath,bbm}
\usepackage{amsthm}
\usepackage{amsmath}
\usepackage{latexsym}
\usepackage{amssymb}
\usepackage{float}
\usepackage{graphicx}           
\usepackage{color}
\usepackage{xcolor}
\usepackage{comment}
\usepackage{enumitem}
\usepackage{multirow}
\usepackage{setspace}
\usepackage[colorlinks=true,linkcolor=blue,citecolor=magenta,urlcolor=blue]{hyperref}
\usepackage{svg}
\usepackage{caption}
\usepackage{subcaption}
\usepackage{tikz}
\usetikzlibrary{shapes.geometric}
\usepackage[normalem]{ulem}

\newcommand{\be}{\begin{equation}}
\newcommand{\ee}{\end{equation}}
\newcommand{\bea}{\begin{eqnarray}}
\newcommand{\eea}{\end{eqnarray}}

\def\squareforqed{\hbox{\rlap{$\sqcap$}$\sqcup$}}
\def\qed{\ifmmode\squareforqed\else{\unskip\nobreak\hfil
\penalty50\hskip1em\null\nobreak\hfil\squareforqed
\parfillskip=0pt\finalhyphendemerits=0\endgraf}\fi}
\def\endenv{\ifmmode\;\else{\unskip\nobreak\hfil
\penalty50\hskip1em\null\nobreak\hfil\;
\parfillskip=0pt\finalhyphendemerits=0\endgraf}\fi}

\newcommand{\tr}{\text{Tr}}
\newcommand{\I}{\mathbbm{1}}
\newcommand{\Q}{\mathcal{Q}}
\newcommand{\C}{\mathcal{C}}

\newcommand{\re}{\color{blue}}  
\newcommand{\blk}{\color{black}}

\makeatletter
\newtheorem*{rep@theorem}{\rep@title}
\newcommand{\newreptheorem}[2]{%
\newenvironment{rep#1}[1]{%
 \def\rep@title{#2 \ref{##1}}%
 \begin{rep@theorem}}%
 {\end{rep@theorem}}}
\makeatother

\newreptheorem{thm}{Theorem}
\newtheorem{lemma}{Lemma}

\begin{document}


\title{Efficient Computation of Generalized Noncontextual Polytopes and Quantum violation of their Facet Inequalities}


\author{Soumyabrata Hazra}
\affiliation{International Institute of Information Technology, Gachibowli, Hyderabad 500032, India}
\author{Debashis Saha}
\affiliation{School of Physics, Indian Institute of Science Education and Research Thiruvananthapuram, Kerala 695551, India}
\affiliation{Department of Physics, School of Basic Sciences, Indian Institute of Technology Bhubaneswar, Odisha 752050, India}
\author{Anubhav Chaturvedi}
\affiliation{Faculty of Applied Physics and Mathematics,
 Gda{\'n}sk University of Technology, Gabriela Narutowicza 11/12, 80-233 Gda{\'n}sk, Poland}
\affiliation{International Centre for Theory of Quantum Technologies (ICTQT), University of Gda{\'n}sk, 80-308 Gda\'nsk, Poland}
\author{Subhankar Bera}
\affiliation{S. N. Bose National Centre for Basic Sciences, Block JD, Sector III, Salt Lake, Kolkata 700106, India}
\affiliation{Department of Physics and Center for Quantum Frontiers of Research and Technology (QFort), National Cheng Kung University, Tainan 701, Taiwan}
\author{A. S. Majumdar}
\affiliation{S. N. Bose National Centre for Basic Sciences, Block JD, Sector III, Salt Lake, Kolkata 700106, India}


\begin{abstract}
Finding a set of empirical criteria fulfilled by any theory satisfying the generalized notion of noncontextuality is a challenging task of both operational and foundational importance. This work presents a methodology for constructing the noncontextual polytope while ensuring that the dimension of the polytope associated with the preparations remains constant regardless of the number of measurements and their outcome size.
The facet inequalities of the noncontextual polytope can thus be obtained in a computationally efficient manner. We illustrate the efficacy of our methodology through several distinct contextuality scenarios. Our investigation uncovers several hitherto unexplored noncontextuality inequalities and demonstrates applications of quantum contextual correlations in certification of non-projective measurements, witnessing the dimension of quantum systems, and randomness certification.
\end{abstract}

\maketitle


\section{Introduction} 

One of the most striking features of quantum theory is that its predictions resist generalized noncontextual or ``Leibnizian" realist explanations \cite{Spekkens2005,Review, mazurek2016experimental,Schmid18,PuseyPRA,XuPRA,Chaturvedi2021characterising, reviewQC}. The notion of noncontextuality embodies the Leibnizian methodological principle \cite{spekkens2019ontologicalidentityempiricalindiscernibles} that attributes identical realist descriptions to operationally equivalent or indistinguishable experimental procedures. The phenomenon of generalized contextuality of quantum theory constitutes a fundamental nonclassical feature of the quantum formalism that underlies other characteristic nonclassical predictions of quantum theory \cite{Spekkens2008prl,Chaturvedi2020quantum,chaturvedi2021quantum,PuseyPRL,LostaglioPRL,SchmidPRX,SchmidGPT20,SchmidFramework,interference,UR,selby2022open,GS,Lostaglio2020contextualadvantage,TavakoliPRXQuantum,mate-arxiv}.
As a rigorous and theory-independent notion of nonclassicality, contextuality serves as a genuine resource enabling quantum-over-classical advantage in a broad range of information processing tasks, such as quantum computation, state discrimination, randomness certification, oblivious communication and
 communication complexity \cite{Spekkens2009,SikoraNJP,howard2014contextuality,Guhne2014,sahaPRA,SahaNJP,Armin2017,Singh2017,PanPRA,Ambainis2019,
SchmidPRL22,State_discri_2,Randomness_certify,gupta2023quantum}.

Preparation noncontextuality attributes identical epistemic states to preparation procedures \cite{Leifer_2014,PhysRevA.75.032110}, which are indistinguishable, i.e., all measurements yield identical statistics on such preparation procedures. Inequalities that hold in theories satisfying preparation noncontextuality can be violated in quantum theory, revealing the contextuality of preparation or simply \emph{preparation contextuality}. 
Similarly, \emph{measurement noncontextuality} attributes identical response schemes to measurement procedures that are operationally indistinguishable, i.e., give rise to identical empirical statistics on all possible preparations. 

Generalized noncontextuality is the logical conjunction of preparation and measurement noncontextuality in scenarios associated with prepare and measure experiments. Analogous to the assumption of local causality that implies Bell inequality, the assumption of generalized noncontextuality implies empirical inequalities, referred to as (generalized) noncontextuality inequalities. Quantum theory prescribes preparations and measurements, which, while satisfying the operational indistinguishable conditions, violate noncontextuality inequalities. 
A contextuality scenario is specified by the number of preparations, measurements, and measurement outcomes, as well as the operational indistinguishability conditions between preparation and measurement procedures corresponding to their distinct convex mixtures, respectively.
Given a contextuality scenario, finding a set of empirical criteria fulfilled by any noncontextual theory is a demanding task of both foundational and operational importance. 

As pointed out by Schmid \textit{et al.} \cite{Schmid18}, the set of empirical statistics possessing noncontextual explanations forms a convex polytope,
and consequently, the inequalities representing the facets of that polytope combine to provide the necessary and sufficient criteria for noncontextuality. To obtain the facet inequalities of the noncontextual polytope, one needs to compute the extremal points of a $D_P-$dimensional polytope associated with the preparations. These extremal points are essentially epistemic states, which are extremal probability distributions over the ontic state space. Subsequently, one computes the extremal points of another polytope associated with measurements. 
Extremal points of the noncontextual polytope are obtained by taking a tensor product of these two polytopes, following a summation over the $\lambda$ variables. It turns out, typically, the dimension of preparation polytope increases exponentially with the number of measurements. 

As mentioned above, the computational technique to retrieve all the facet inequalities applicable to arbitrary contextuality scenarios is computationally challenging. Therefore, it is highly desirable to seek efficient methods to find a set of empirical conditions depicting the generalized non-contextuality framework. In the present study, our aim is to efficiently formulate the noncontextual inequalities.

Specifically, here we introduce a novel method to retrieve noncontextuality inequalities in any contextuality scenario, where only a single ontic state is needed to characterize the polytope for preparations. As a result, in contrast to the conventional method \cite{Schmid18}, in our approach, one needs to compute the extremal points of preparation polytope whose dimension remains constant, irrespective of the number of measurements and their outcomes. The formalism proposed here enables us to obtain the noncontextual polytope considerably faster. The facet inequalities of this polytope constitute noncontextuality inequalities. Violation of the obtained inequalities thus provides us with necessary and sufficient conditions
for guaranteeing generalized quantum contextual correlations.

As an upshot of our formalism, through the present analysis, we are able to investigate efficiently various contextuality scenarios and explore quantum violation of noncontextuality inequalities, yielding several insights into the different ways Nature exhibits contextuality. 
As a result we uncover several new applications of quantum contextuality in those scenarios, such as certification of non-projective measurements, certification of dimensionality, quantum advantage in oblivious communication, and randomness certification.

The problem of whether a given numerical dataset of outcome probabilities admits a noncontextual explanation can be answered via a linear program using Farkas lemma \cite{andersen2001certificates, luenberger1984linear}. However, without explicit noncontextual inequalities there is no structured way to construct quantum contextual behaviors.
It is noteworthy that rigorous studies have been limited to only two contextuality scenarios so far. These include the simplest contextuality scenario, which involves four preparations and two measurements with indistinguishability conditions for preparation \cite{PuseyPRA,Schmid18,Lostaglio2020contextualadvantage,Wagner_2024}, as well as the scenario featuring six preparations and three measurements with indistinguishability conditions for both preparations and measurements \cite{Schmid18}. Another scenario involving six preparations and three binary outcome measurements with indistinguishability condition only on preparations has been studied in the context of witnessing coherence in a Mach-Zehnder interferometer \cite{Wagner2024coherence,doi:10.1126/sciadv.adj4249}. In light of this, an additional objective of this work is to thoroughly investigate other elementary contextuality scenarios and provide a comprehensive analysis. Explicit characterizations of these prepare and measure scenarios also enable future study of prepare-transform-measure scenarios and investigate the role of transformations in contextual advantages. The collection of enlisted noncontextual inequalities across scenarios may in future be used as training data for machine learning models to uncover novel structures or hidden symmetries within the polytopes.

The rest of the paper is organized as follows. Section II provides a comprehensive overview of the generalized notion of contextuality scenarios. In the subsequent section, we present our approach to obtain the set of necessary and sufficient conditions for noncontextuality in an arbitrary contextuality scenario. Next, in the simplest contextuality scenario, we rigorously demonstrate our methodology. Next, we describe our method explicitly stating the algorithm for obtaining noncontextuality inequalities and discuss its merits in contrast to the standard approach. The following section investigates different contextuality scenarios using the proposed methodology. We consider six scenarios with indistinguishability conditions only among preparations, as well as two additional scenarios with indistinguishability conditions among both preparations and measurements. These scenarios encompass a range of four to nine preparations and a maximum of three measurements. Consequently, we retrieve a large number of novel noncontextuality inequalities in these scenarios. To obtain the maximum quantum violations of these noncontextuality inequalities, we employ two techniques of semi-definite programming introduced in \cite{Chaturvedi2021characterising}. The see-saw technique retrieves lower bounds on the maximum quantum violations along with the quantum states and measurements. On the other hand, the second technique, inspired by the Navascu\'es--Pironio--Ac\'in hierarchy for nonlocal correlations \cite{npa}, provides a dimension-independent upper bound on the maximum quantum violation of the noncontextuality inequalities. We also study the robustness of the experimental noise of the quantum violations. In Section V, we delve into various intriguing applications stemming from quantum violations of our discovered noncontextuality inequalities. We demonstrate that the quantum violation of some of these inequalities can serve to certify the dimensionality of quantum systems, non-projective measurements, and quantum randomness. Finally, we conclude in Section VI with a summary of our results and outline promising directions for further studies.

\section{Generalized notion of contextuality} 

Consider, a prepare and measure experiment entailing several distinct preparation and measurement procedures. A preparation procedure is labelled by $P_x$, where $x$ denotes the specific preparation, and a measurement procedure is denoted by $M_{z|y}$, where $z$ and $y$ represent the outcome and setting of the measurement, respectively. Using an operational theory, such as quantum theory, we can make predictions about the empirical statistics $\{p(z|x,y)\}$, where $p(z|x,y)$ indicates the probability of obtaining outcome $z$ when the measurement specified by $y$ is performed on the preparation specified by $x$. 
We say that two preparation procedures, $P_x$ and $P_{x'}$, are operationally equivalent or indistinguishable (denoted as $P_x \sim P_{x'}$) if they yield identical outcome statistics for all measurements,
\be 
p(z|x,y)=p(z|x',y), \   \forall M_{z|y}. 
\ee   
Similarly, two measurement procedures $M_{z|y}$ and $M_{z'|y'}$ are operationally equivalent or indistinguishable (denoted as $M_{z|y} \sim M_{z'|y'}$), if they produce identical outcome statistics for all possible preparations 
\be 
p(z|x,y)=p(z'|x,y'), \  \forall P_x. 
\ee

Let us consider a prepare and measure experiment involving $n_x$ distinct preparations as $x \in \{0,\dots, n_x-1\}$ and $n_y$ different measurements as $y\in \{0,\dots, n_y-1\}$, with each measurement having $n_z$ possible outcomes as $z\in \{ 0,\dots,n_z-1 \}$. In this experiment, we have a set of hypothetical preparations, each of which is realized by taking convex mixtures of these $n_x$ preparations such that the resultant mixed preparations are indistinguishable. These mixed preparations are labeled by the variable $s\in \{0,\dots,n_s\}$ and they are realized by the set of convex coefficients $\{\alpha_{x|s}\}$, satisfying
\be 
\alpha_{x|s} \geqslant 0, \ \  \sum_{x} \alpha_{x|s}=1, \forall x,s . 
\ee
The indistinguishability conditions imply that for all $ s,s' \in \{0,\dots,n_s\}$,
\be \label{ecp}
\sum_x \alpha_{x|s} P_x  \sim  \sum_x \alpha_{x|s'} P_x .
\ee 
It is important to note that the above set of indistinguishability conditions are taken to be linearly \textit{independent}, meaning that no indistinguishability condition can be deduced from the other conditions. Mathematically, this requires the vectors $\{\vec{u}_s\}_s$ of these convex coefficients,
\[ 
\vec{u}_s = \left( \alpha_{0|s}, \alpha_{1|s}, \cdots, \alpha_{n_x-1|s} \right)
\]
to be a linearly independent set of vectors.

Similarly, we have indistinguishable measurement procedures labelled by $t \in \{0,\dots, n_t\}$, each of them is realized by different taking convex mixtures with the coefficients $\{\beta_{z,y|t}\}$ satisfying
\be 
\beta_{z,y|t}  \geqslant 0, \ \sum_{z,y|t} \beta_{z,y|t}=1, \ \forall z,y,t.
\ee
Thereupon, these indistinguishability conditions on measurements can be expressed as 
\be \label{ecm}
\sum_{z,y} \beta_{z,y|t} M_{z|y} \sim \sum_{z,y} \beta_{z,y|t'} M_{z|y} \ , 
\ee 
for all $t,t' \in \{0,\dots,n_t\}$.
Here also, we consider these indistinguishability conditions to be linearly independent, which requires the vectors 
\[ 
\vec{v}_t = \left( \beta_{0|t}, \beta_{1|t}, \cdots, \beta_{n_y-1|t} \right)
\]
to form a linearly independent set of vectors.
The number of preparations, measurements, and measurement outcomes, together with the set of linearly independent indistinguishability conditions, defines a \textit{contextuality scenario} uniquely, upon considering all possible symmetries of involved operational quantities. As an example, consider two contextuality scenarios with indistinguishability condition on preparation being defined as \begin{align}
    \alpha_i P_0+(1-\alpha_i) P_1\sim\alpha_j P_2+(1-\alpha_j)P_3
\nonumber\end{align} and \begin{align}
    \alpha_i P_0+(1-\alpha_i) P_2\sim\alpha_j P_1+(1-\alpha_j)P_3\nonumber
\end{align} respectively, where $0<\alpha_i,\alpha_j<1$. These two scenarios are exactly equal, upon relabeling $P_1$ by $P_2$ and vice-versa. More on possible symmetry operations is discussed in Section \ref{sec:method}.

In quantum theory, preparations are described by density operators $\rho_x$, and measurements are described by positive semi-definite operators $\mathbbm{M}_{z|y}$, satisfying $\sum_z \mathbbm{M}_{z|y}=\I$, where $\I$ is the identity operator. The probability of obtaining outcome $z$ when performing measurement $M_{z|y}$ on preparation $P_x$ is given by $p(z|x,y) = \text{Tr}(\rho_x \mathbbm{M}_{z|y})$. 
Furthermore, quantum preparations and measurements satisfy the indistinguishability conditions given by \eqref{ecp} and \eqref{ecm} if and only if the following equalities hold:
\be 
\sum_x \alpha_{x|s} \ \rho_x  = \sum_x \alpha_{x|s'} \ \rho_x \ , \ \ \forall s,s' \in \{0,\dots,n_s\}.
\ee 
and 
\be 
\sum_{z,y} \beta_{z,y|t} \mathbbm{M}_{z|y} = \sum_{z,y} \beta_{z,y|t'} \mathbbm{M}_{z|y} \ , \ \forall t,t' \in \{0,\dots,n_t\} .
\ee 

An ontological model \cite{Harrigan_2010} offers an explanation to the prediction of an operational theory by considering the state of the system to be an objective reality. This state is called the \textit{ontic state} of the system, denoted by $\lambda \in \Lambda$, where $\Lambda$ is an arbitrary measurable space referred to as the ontic state space. A preparation procedure $P_x$ prepares the system in an ontic state $\lambda$ with probability $\mu(\lambda|x)$. The distribution $\mu(\lambda|x)$ is referred to as the epistemic state of the system. On the measurement side, the probability of obtaining the outcome $z$ when a measurement $M_{z|y}$ is performed on the ontic state $\lambda$ is given by the response function $\xi(z|\lambda,y)$. An ontological model satisfying the generalized notion of noncontextuality assigns identical epistemic states to indistinguishable mixed preparations and identical response functions to indistinguishable mixed measurement procedures \cite{Spekkens2005}.
More precisely, the indistinguishability conditions \eqref{ecp} and \eqref{ecm} in any noncontextual ontological model imply
\be \label{ncpc}
\sum_x \alpha_{x|s} \mu(\lambda|x) = \sum_x \alpha_{x|s'} \mu(\lambda|x) ,
\ee 
for all $s,s' \in \{0,\dots,n_s\}$  and 
\be \label{ncmc}
 \sum_{z,y} \beta_{z,y|t} \xi(z|\lambda,y) = \sum_{z,y} \beta_{z,y|t'} \xi(z|\lambda,y) ,
\ee 
for all $t,t'\in \{0,\dots,n_t\}$, pertaining to every $\lambda$. An operational theory is called noncontextual if its predictions can be explained by a noncontextual ontological model.
Given a contextuality scenario, the set of empirical probabilities $\{p(z|x,y)\}$ obtained from any noncontextual operational theory forms a polytope \cite{Schmid18}. 
A general form of a facet inequality of the noncontextual polytope is given by
\be \label{gfi}
\sum_{x,y,z} c_{x,y,z} \ p(z|x,y) \leqslant \C ,
\ee 
where $c_{x,y,z}$ are real coefficients and $\C$ is the noncontextual bound. An operational theory whose predictions violate such inequalities is said to be contextual. There exists $\{p(z|x,y)\}$ predicted by quantum theory that violates such inequalities in general.

\section{Construction of the noncontextual polytope }\label{sec:method}

This section presents details of our approach to derive noncontextuality inequalities for any given contextuality scenario. The fundamental underpinning concept entails retrieving the extremal points of a polytope that characterizes the epistemic states corresponding to the preparations in an ontological model. Crucially, these extremal points are formulated to be independent of the number of ontic states while encompassing all feasible response functions describing the measurements. 

\textit{\textbf{Polytope characterizing preparations.}} Let us recall that the presence of indistinguishability conditions in a given contextuality scenario enforces the relationship \eqref{ncpc} on the epistemic states ${\mu(\lambda|x)}$, where ontic state $\lambda$ can take arbitrary possible values. Let us now introduce the following additional variable,
\be \label{defqx}
q(x,\lambda) = \frac{\mu(\lambda|x)}{\sum_x \alpha_{x|s} \mu(\lambda|x)}  =: \frac{\mu(\lambda|x)}{\overline{\mu}(\lambda)} .
\ee 
Note that the denominator of the above expression is a constant for all $s$, and has the expression
\be \label{omul}
\overline{\mu}(\lambda) = \sum_x \alpha_{x|s} \mu(\lambda|x) .
\ee 
Also, $\overline{\mu}(\lambda)> 0$ for all $\lambda$.
By dividing both sides of Eq. \eqref{ncpc} by $\overline{\mu}(\lambda)$, we find that our new variables $\{q(x,\lambda)\}$ satisfy the following conditions for all $\lambda$,
\be \label{ncpcq}
\forall s, \ \sum_x \alpha_{x|s} q(x,\lambda) = 1 .
\ee 
Hence, for each $\lambda$, the collection of variables $\{q(x,\lambda)\}_x$ constitutes a convex polytope that adheres to the positivity constraint $q(x,\lambda) \geqslant 0$, as well as to the constraints implied by Eq. \eqref{ncpcq}. Let's denote the extremal points of this polytope as $e_p$, and the corresponding values at these extremal points as $q(x|e_p)$. In simpler terms, any $q(x,\lambda)$ can be expressed in the following manner:
\be \label{qxl}
q(x,\lambda) = \sum_{e_p} w(e_p|\lambda) q(x|e_p) ,
\ee 
where $w(e_p|\lambda)$ are convex weights relative to each specific $\lambda$.\\

\textit{\textbf{Polytope characterizing measurements.}} 
On the other side, for each $\lambda$, the set of quantities $\{\xi(z|\lambda,y)\}_{z,y}$ forms a convex polytope that fulfills the criteria of positivity ($\xi(z|\lambda,y)\geqslant 0$), normalization ($\sum_z \xi(z|\lambda,y) = 1$), and adheres to the constraints dictated by the indistinguishability condition in Eq. \eqref{ncmc}. We assign the label $e_m$ to the extremal points of this polytope, with the corresponding value at these points denoted by $\xi(z|y,e_m)$. Consequently, any $\xi(z|\lambda,y)$ can be represented in the following way
\be \label{xizy}
\xi(z|\lambda,y) = \sum_{e_m} w(e_m|\lambda) \xi(z|y,e_m) .
\ee 
Here, $w(e_m|\lambda)$ constitutes a valid set of convex weights pertaining to each $\lambda$. \\

\begin{figure}[!ht]
    \centering
    \includegraphics[scale=0.45]{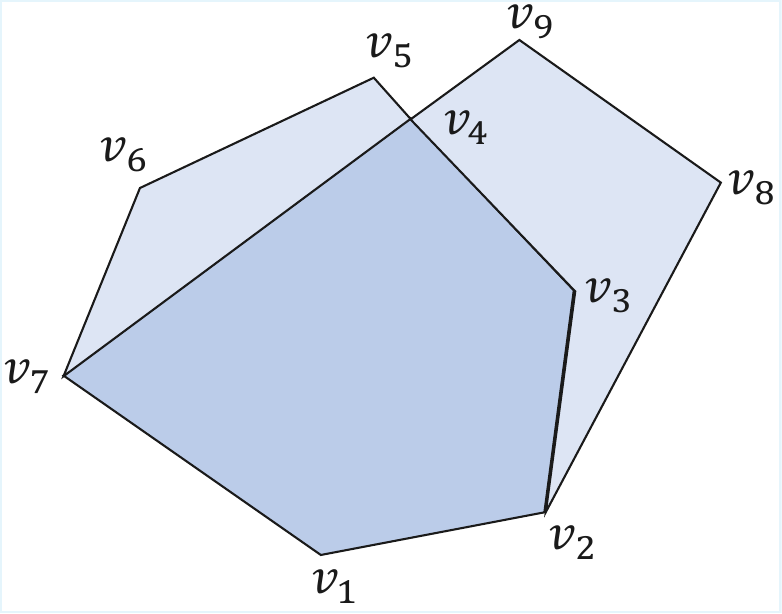}
    \caption{ The extended polytope $(\mathbb{P}_\text{P})$ encompassing the probabilities specified by \eqref{pbar}, is defined by the collection of vertices as, $\mathbb{P}_\text{P}=\{v_1,v_2,v_3,v_5,v_6,v_7\}$. The polytope adhering solely to the normalization condition \eqref{n}, is defined as, $\mathbb{P}_{\text{NP}}=\{v_1,v_2,v_7,v_8,v_9\}$. $\mathbb{P}_{\text{NCP}}=\{v_1,v_2,v_3,v_4,v_7\} $ is the precise non-contextual polytope, which is the intersection of $\mathbb{P}_\text{P}$ and $\mathbb{P}_\text{NP}$.}
    \label{fig:polyope}
\end{figure}

\textit{\textbf{The combined polytope.}} We now articulate the empirical probability $p_{NC}(z|x,y)$ stemming from non-contextual operational theories in terms of the extremal distributions $q(x|e_p)$ and $\xi(z|y,e_m)$ using following sequence of mathematical relations:
\begin{widetext}
\bea  \label{simpzxy}
p_{NC}(z|x,y) &=& \int_\lambda \xi(z|\lambda,y) \mu(\lambda|x) \ d\lambda \label{ncp probabilities} \\
&=& \int_\lambda \sum_{e_m} w(e_m|\lambda) \xi(z|y,e_m) \mu(\lambda|x) \ d\lambda \nonumber \\ 
&=& \sum_{e_m} \xi(z|y,e_m) \left( \int_\lambda  w(e_m|\lambda) \mu(\lambda|x) \ d\lambda  \right) \nonumber \\
& = & \sum_{e_m} \xi(z|y,e_m) \left( \int_\lambda \overline{\mu}(\lambda) w(e_m|\lambda) q(x,\lambda)  d\lambda \right) \nonumber \\
& = & \sum_{e_m} \xi(z|y,e_m) \left( \int_\lambda \overline{\mu}(\lambda) w(e_m|\lambda) \left( \sum_{e_p} w(e_p|\lambda) q(x|e_p) \right)  d\lambda \right) \nonumber \\
& = & \sum_{e_p,e_m} q(x|e_p) \xi(z|y,e_m) w(e_p,e_m) .
\label{ncp_probabilities_end}\eea  
\end{widetext}
The second, fourth, and fifth lines can be deduced from their respective preceding lines through the application of Eqs. \eqref{xizy}, \eqref{defqx}, and \eqref{qxl}, respectively. The third and sixth lines involve a rearrangement of the summation terms, where we use
\be 
 w(e_p,e_m):= \int_\lambda \overline{\mu}(\lambda) w(e_p|\lambda)w(e_m|\lambda) ~d\lambda.
\label{convex_weight}\ee  
Now, $\{\omega(e_p,e_m)\}$ forms a set of convex weights since they satisfy $w(e_p,e_m) \geqslant 0$ and \be\sum_{e_p,e_m} w(e_p,e_m) = \int_\lambda \overline{\mu}(\lambda) d\lambda = 1.\ee
The second equality follows by summing over $\lambda$ on both sides of Eq. \eqref{omul}.
So, the probabilities $p_{NC}(z|x,y)$ in any non-contextual theory are necessarily confined within the polytope whose extremal points emerge from the multiplication of $e_p$ and $e_m$. In other words, the probabilities $p_{NC}(z|x,y)$ lie within the polytope determined by the extremal probability distributions $\{q(x|e_p) \xi(z|y,e_m)\}$, i.e., $\mathbb{P}_{\text{NCP}}\subseteq \mathbb{P}_{\text{P}}$ (see Fig. \ref{fig:polyope}). In general, an interior point of this bigger polytope $\mathbb{P}_{\text{P}}$ can be written as 
\be \label{pbar}
\widetilde{p}(z|x,y) = \sum_{e_p,e_m} \nu_{e_p,e_m} q(x|e_p) \xi(z|y,e_m), 
\ee where $\{\nu_{e_p,e_m}\}$ is some set of convex weights. Since an interior point of a polytope can have several distinct convex decompositions, $\{\nu(e_p,e_m)\}$ cannot be expressed as a product of two independent set convex weights associated with $e_p$ and $e_m$, in general.  
As the extremal points of the extended polytope $\mathbb{P}_{\text{P}}$ are independent of $\lambda$, only a single ontic state is required to characterize this polytope.
We can reconfirm, using the above relation, that $\widetilde{p}(z|x,y)$ indeed satisfy the indistinguishability conditions \eqref{ecp} since
\begin{eqnarray} 
&& \sum_x \alpha_{x|s} \widetilde{p} (z|x,y) \nonumber \\ 
& = & \sum_{e_p,e_m} \nu_{e_p,e_m} \left( \sum_x \alpha_{x|s} q(x|e_p) \right) \xi(z|y,e_m) \nonumber \\
& = & \sum_{e_p,e_m} \nu_{e_p,e_m}\xi(z|y,e_m)
\end{eqnarray} remains constant for all $s$, in accordance with \eqref{ncpcq}. Similarly, $\widetilde{p}(z|x,y)$ also adheres to the indistinguishability condition \eqref{ecm}.
However, the expression
\be \label{sumpbar}
\sum_z \widetilde{p}(z|x,y) = \sum_{e_p,e_m}\nu_{e_p,e_m} q(x|e_p)
\ee
does not necessarily equate to $1$, thus violating the normalization condition. This implies $\mathbb{P}_{\text{NCP}} \subset \mathbb{P}_\text{P}$ (refer to Fig. \ref{fig:polyope}) Nevertheless, all observed probabilities should satisfy the normalization requirement:  
\be \label{n}
\sum_z \widetilde{p}(z|x,y)=1,
\ee 
which in turn implies \be \label{norm_cond}\sum_{e_p,e_m} \nu_{e_p,e_m} q(x|e_p)=1 .\ee In fact, there exists a polytope wherein the probabilities conform solely to the normalization conditions \eqref{n}. To rectify this issue with the extended polytope, our strategy is to identify the facet inequalities of the polytope defined by \eqref{pbar} while it is constrained by the polytope where probabilities adhere to the normalization conditions \eqref{n}. The actual noncontextual polytope is essentially the intersection of the extended polytope and the normalization polytope. For a visual representation, refer to Fig. \ref{fig:polyope}.  We now present this as a formal proof in the following lemma. \begin{lemma} The intersection of extended polytope $\mathbb{P}_{\text{P}}$ and the normalization polytope $\mathbb{P}_\text{NP}$ is exactly equal to the noncontextual polytope $\mathbb{P}_\text{NCP}$. \begin{proof}
     Following our earlier discussion from Eq. \eqref{ncp_probabilities_end} to Eq. \eqref{sumpbar}, we have already shown that $\mathbb{P}_\text{NCP}$ is a strict subset of $\mathbb{P}_\text{P}$. It is also not difficult to see that $\mathbb{P}_\text{NCP}\subset \mathbb{P}_\text{NP}$. If we can show that any interior point of $\mathbb{P}_\text{P}\cap \mathbb{P}_\text{NP}$ also belongs to $\mathbb{P}_\text{NCP}$, then the desired result is established. Any interior point of $\mathbb{P}_\text{P}\cap \mathbb{P}_\text{NP}$ satisfies Eq. \eqref{pbar} and \eqref{norm_cond}. The noncontextual probabilities $p_{NC}(z|x,y)$ satisfy \eqref{ncp_probabilities_end}, where $w(e_p,e_m)$ plays the same role as $\nu(e_p,e_m)$ in \eqref{pbar} and also \be \sum_{e_p,e_m}w(e_p,e_m)q(x|e_p) = 1, \ee as they are valid probabilities. This implies $\mathbb{P}_\text{NCP}\subseteq \mathbb{P}_\text{P}\cap \mathbb{P}_\text{NP}$, i.e., there can still be some interior point of $\mathbb{P}_\text{P}\cap \mathbb{P}_\text{NP}$ for which $\nu_{e_p,e_m}$ may not satisfy Eq. \eqref{convex_weight}. However, as we have argued earlier, $\nu_{e_p,e_m}$ may not be factorizable in terms of two independent sets of convex weights associated with $e_p$ and $e_m$. Therefore, a general expression can always be written as \be \nu_{e_p,e_m} = \int_{\lambda} w(e_p|\lambda)w(e_m|\lambda) \gamma(\lambda)~ d\lambda ,\ee where $\{\gamma(\lambda)\}$ 
    is a probability distribution over ontic space, that is independent of the preparation variable $x$. 
     Considering the existence of a noncontextual ontological model, the only way of realizing $\gamma(\lambda)$, is to take a 
    convex mixture of epistemic states corresponding to an indistinguishability relation, i.e., \be \gamma(\lambda)=\sum_x\alpha_{x|s}\mu(\lambda|x)=\overline{\mu}(\lambda) \ee for all $\lambda$. 
    So, we can write, 
    \be \nu_{e_p,e_m}=\int_\lambda w(e_p|\lambda)w(e_m|\lambda)\sum_x\alpha_{x|s}\mu(\lambda|x)~d\lambda. \label{epemproof}\ee 
    Finally, using this expression in \eqref{pbar}, we retrieve the exact same form of $\widetilde{p}(z|x,y)$ as for $p_{NC}(z|x,y)$. This concludes the proof.  
\end{proof}   
\end{lemma}  Prior to presenting the succinct algorithm outlining our method to derive noncontextuality inequalities, we first demonstrate the method by explicitly applying it to the simplest contextuality scenario.

\textit{\textbf{An explicit example.}}
The simplest contextuality scenario involves four preparations indexed by $x\in \{0,1,2,3\}$, such that they obey the following indistinguishability condition
\be \label{ec21porac}
\frac{1}{2} \left( P_0 + P_1 \right) \sim \frac{1}{2} \left( P_2 + P_3 \right) .
\ee 
Furthermore, there are two measurements, indexed by $y \in \{0,1\}$, each yielding binary outcomes $z \in \{0,1\}$. No non-trivial indistinguishable conditions are applied to the measurements in this contextuality scenario.
According to \eqref{ncpcq}, the variables $\{q(x)\}$ describing the preparations adhere to the relations
\be 
q(x) \geqslant 0, \ q(0) + q(1) = q(2) + q(3) = 2,
\ee 
while the variables $\{\xi(z|y)\}$ characterizing the measurements satisfy
\be 
\xi(z|y) \geqslant 0 , \ \xi(0|0) + \xi(1|0) = \xi(0|1) + \xi(1|1) =1 .
\ee 
Clearly, the set of allowed values of these variables form convex polytopes. It is easy to see that there are four extremal points of both these polytopes which are labeled as $e_p, e_m \in \{1,2,3,4\}$,
\begin{table}[!ht]
    \centering
    \begin{tabular}{c|c c c c}
    \hline 
      $e_p$ & $q(0|e_p)$ & $q(1|e_p)$ & $q(2|e_p)$ & $q(3|e_p)$ \\
         \hline 
         1 & 2 & 0 & 2 & 0 \\
         \hline
         2 & 2 & 0 & 0 & 2 \\
       \hline 
        3 & 0 & 2 & 2 & 0 \\
       \hline 
        4 & 0 & 2 & 0 & 2 \\
       \hline 
    \end{tabular} 
\end{table} 
\begin{table}[!ht]
    \centering
    \begin{tabular}{c|c c c c}
    \hline 
      $e_m$ & $\xi(0|0,e_m)$ & $\xi(1|0,e_m)$ & $\xi(0|1,e_m)$ & $\xi(1|1,e_m)$ \\
         \hline 
         1 & 1 & 0 & 1 & 0 \\
         \hline
         2 & 1 & 0 & 0 & 1 \\
       \hline 
        3 & 0 & 1 & 1 & 0 \\
       \hline 
        4 & 0 & 1 & 0 & 1 \\
       \hline 
    \end{tabular} 
\end{table} \\

By taking the tensor product of these extremal points, we obtain $16$ extremal distributions $\{q(x|e_p) \xi(z|y,e_m)\}$, each corresponding to a specific pair $(e_p,e_m)$. 
Subsequently, we extract the facet inequalities of the polytope defined by these $16$ extremal distributions, resulting in a total of $24$ facet inequalities for this case.
First, by utilizing the indistinguishable condition \eqref{ec21porac},
\be 
\forall y, \ p(z|3,y) =  p(z|0,y) + p(z|1,y) - p(z|2,y) ,
\ee 
we can substitute the probabilities $\{p(z|3,y)\}$ in the facet inequalities to make them simpler.  Secondly, the polytopes of variables $\{q(x)\}$ and $\{\xi(z|y)\}$ satisfy symmetry conditions with respect to interchanging variables. For example, if variables $x=0$ and $x=1$ are interchanged, then the four extremal points $\{q(x|e_p)\}$ remain unchanged, and consequently, the associated polytope retains the same form. Let us denote such symmetry by $P_0 \longleftrightarrow P_1$. Similarly, we can identify all the symmetries in this scenario 
\bea \label{s1sym}
& P_0 \longleftrightarrow P_1 \nonumber ,\\ 
& P_2 \longleftrightarrow P_3 \nonumber , \\
& M_{0|y} \longleftrightarrow M_{1|y} \quad \forall y=0,1 \nonumber  ,\\ 
& M_{z|0} \longleftrightarrow M_{z|1} \quad \forall z=0,1 .
\eea  
When an inequality becomes identical to another after applying any of the symmetries, it indicates that these two inequalities are equivalent. After systematically applying all of the symmetries \eqref{s1sym}, we can identify a set of inequalities that are equivalent to each other and form an equivalence class. It is sufficient to consider any one representative inequality from that equivalent set or equivalent class of inequalities. The `orbit size' signifies the number of inequalities within a specific equivalence class. Among the $24$ inequalities initially obtained, the different in-equivalent classes of facet inequalities are enlisted below, where we have used a short-hand notation $p_{x,y}^z := p(z|x,y)$.
\begin{table}[!ht]
    \centering
    \begin{tabular}{c|c}
    \hline 
      orbit & Inequalities \\ size &  \\
         \hline 
         \hline 
         9 & $ - p^1_{2,0} \leqslant 0$  \\
       \hline 
       3 & $p^0_{0,0} + p^0_{1,0} + p^1_{2,0} \leqslant 2$  \\
       \hline 
       2 & $- p^0_{0,0} - p^1_{0,0} + p^1_{0,1} \leqslant 0$   \\
       \hline 
        1 & $p^0_{0,0} + p^1_{0,0} + p^1_{1,1} \leqslant 2$   \\
       \hline 
        3 & $p^0_{1,0} + p^1_{2,0} + p^1_{0,1} - p^1_{2,1} \leqslant 2$   \\
       \hline 
       3 & $-p^0_{2,1} - p^1_{0,0} + p^1_{2,0} - p^1_{1,1} \leqslant 0$   \\
       \hline 
        1 & $-p^0_{0,0} - p^0_{1,0} + p^0_{2,1} - p^1_{2,0} + p^1_{2,1} \leqslant 0$   \\
       \hline 
       1 & $p^0_{0,0} + p^0_{1,0} - p^0_{2,1} + p^1_{0,0} + p^1_{2,0} - p^1_{0,1} \leqslant 0$   \\
       \hline 
       1 & $-p^0_{0,0} - p^0_{1,0} + p^0_{2,1} - p^1_{0,0} - p^1_{2,0} + p^1_{0,1} \leqslant 0$   \\
       \hline 
    \end{tabular}
\end{table} 

Some of these inequalities are violated by quantum theory. It is apparent that the normalization condition does not hold since the quantity \eqref{sumpbar} is not $1$ for any of the extremal points of $e_p$ given above.
Therefore, we will consider the facet inequalities of the polytope intersected with the polytope where the probabilities satisfy the normalization conditions \eqref{n}. This can be readily done by substituting $p(1|x,y)$ in the facet inequalities using the normalization condition 
\be 
p(1|x,y) = 1 - p(0|x,y) \ \forall x,y.
\ee 
Further, imposing the symmetries \eqref{s1sym}, we find that the number of in-equivalent classes reduces to only two:
\begin{table}[!ht]
    \centering
    \begin{tabular}{c|c}
    \hline 
      orbit size & Inequalities \\
         \hline 
         \hline 
         16 & $ p^0_{2,0} \leqslant 1$  \\
       \hline 
         8 & $p^0_{1,0} - p^0_{2,0} - p^0_{0,1} + p^0_{2,1} \leqslant 1$  \\
       \hline 
    \end{tabular}
    \caption{We obtained $24$ inequalities for the simplest contextuality scenario described in \eqref{ec21porac}. This set of inequalities reduces to two in-equivalent classes after applying the indistinguishable conditions, symmetries, and normalization conditions.}
    \label{table-scenario1}
\end{table} 

The inequalities of Table \ref{table-scenario1} have already been reported in earlier works \cite{Schmid18, PuseyPRA}. While the first inequality is trivial, the second inequality is violated in quantum theory, achieving the maximum value $\sqrt{2}\approx 1.414$. It is interesting to note that this second inequality is identical to the success metric of the parity oblivious random access codes \cite{Spekkens2009}.

\re 
\blk 
\subsection*{Algorithm to obtain the set of noncontextuality inequalities}

Now, we summarize our algorithm, outlining the step-by-step process for deriving noncontextuality inequalities for any prepare and measure contextuality scenario.

\begin{itemize}
\item First obtain the extremal points $\{q(x|e_p)\}$ of the polytope for the variables $\{q(x)\}$ satisfying the conditions:
\begin{align} 
 & (i) \ q(x) \geqslant 0 ; \label{p1}\\
& (ii) \ \forall s, \ \sum_x \alpha_{x|s} q(x) = 1 . \label{p2}
\end{align} 
\item Next, evaluate the extremal points $\{\xi(z|y,e_m)\}$ of the polytope of the variables $\{\xi(z|y)\}$ satisfying the conditions:
\begin{align}
& (i) \ \xi(z|y) \geqslant 0 ; \label{m1}\\
& (ii) \  \forall y, \ \sum_z \xi(z|y) = 1; \label{m2} \\
& (iii) \ \forall t,t', \ \sum_{z,y} \beta_{z,y|t} \xi(z|y) = \sum_{z,y} \beta_{z,y|t'} \xi(z|y)  . \label{m3} 
\end{align} 
In the case of only indistinguishability conditions on preparations, the polytope is obtained using conditions $(i)$ and $(ii)$ only.
\item Multiply the extremal points obtained in the previous steps to yield the extremal points of the extended polytope $(\mathbb{P}_{\text{P}})$ as ${q(x|e_p)\cdot \xi(z|y,e_m)}$. Subsequently, determine the facet inequalities of the extended polytope. 
\item Impose the normalization condition \eqref{n} to the derived facet inequalities. For example, $p(z=k-1|x,y)$ can be replaced by $1-\sum_{z=0}^{k-2} p(z|x,y)$ for all $x,y$, so that the revised facet inequalities do not contain $p(k-1|x,y)$. Further reduce number of probabilities $\{p(z|x,y)\}$ in the facet inequalities using the indistinguishability conditions \eqref{ecp} and \eqref{ecm},
\be \label{ecpp}
\forall s,s', \forall z,y, \ \sum_x \alpha_{x|s} p(z|x,y)  =  \sum_x \alpha_{x|s'} p(z|x,y) ,
\ee
\be \label{ecmp}
\forall t,t', \forall x, \ \sum_{z,y} \beta_{z,y|t} p(z|x,y) = \sum_{z,y} \beta_{z,y|t'} p(z|x,y) .
\ee
\end{itemize}
 Finally, identify the symmetries on the variables $x,y,z,$ such that the extremal points $\{q(x|e_p)\cdot \xi(z|y,e_m)\}$ remain invariant, and apply these symmetries to the revised facet inequalities obtained in the previous step in order to obtain distinct in-equivalent classes of noncontextuality inequalities.

\subsection*{Comparison with the method of \cite{Schmid18}  } 
Method to find the noncontextual polytope was initially provided by Schmid {\it et al.} \cite{Schmid18}. However, our approach presents a two-folded and substantial advantage over this method in terms of efficiency.

Recall that $n_x,n_y$ and $n_z$ refer to the number of preparations, measurements and outcomes, respectively, in a contextuality scenario. Say, the number of extremal points obtained for the variables $\{\xi(z|y)\}$ satisfying \eqref{m1}-\eqref{m3} is $r$. Following the method in \cite{Schmid18}, the total number of variables $\{\mu(\lambda|x)\}$ sufficient to characterize the polytope associated with preparation is $rn_x $. However, owing to the normalization conditions and the independent indistinguishability conditions \eqref{ncpc}, $n_x$ and $r\cdot n_s$ number of variables are eliminated, respectively. As a result, the dimension of the polytope characterizing the preparations $\{\mu(\lambda|x)\}$, becomes \be D_P = (n_x-n_s)r - n_x  \nonumber\ee \cite{polytope}.  The widely used Avis–Fukuda reverse-search algorithm \cite{AvisFukuda1996}, which enumerates the vertices of a polytope defined by linear constraints, attains linear time complexity with respect to the dimension, for a constant number of vertices, i.e., $\mathcal{O}(D_P) = \mathcal{O}(r(n_x-n_s))$ for the preparation polytope. Note that $r$ typically grows exponentially with the number of measurements $n_y$; for instance, when no indistinguishability conditions are imposed on the measurements, we have $r = n_z^{n_y}$. Thus, the time complexity of the standard approach \cite{Schmid18} to vertex enumeration is exponential in $n_y$. In contrast, our method involves a fixed number of independent variables for characterizing the preparations irrespective of the settings in the measurement side, which is \be n_x-n_s \nonumber\ee (see \eqref{p1}-\eqref{p2}) . Consequently, the reverse-search algorithm adapting to our framework achieves a time complexity of $\mathcal{O}(n_x-n_s)$. Hence, the proposed method enjoys an exponential advantage in $n_y$ for \textit{vertex enumeration problem} on the preparation side.


A second advantage arises in enumerating the vertices of the noncontextual polytope. The standard approach \cite{Schmid18}, requires solving the vertex enumeration problem for two distinct polytopes corresponding to the variables $\{\mu(\lambda|x)\}$ and $\{\xi(z|y,\lambda)\}$. Then one needs to compute the tensor product polytope and subsequently, obtain the projection polytope of the tensor product polytope by summing over the variables $\lambda$, to get the actual noncontextual polytope $\mathbbm{P}_{NCP}$. It involves $r\cdot n_y\cdot n_z$ number of variables $\{\xi(z|y,\lambda)\}$ that describe the measurements. And, owing to the normalization conditions and the independent indistinguishability conditions \eqref{ncmc}, we can eliminate $r\cdot n_y$ and $r\cdot n_t$ number of variables, respectively. As a result, the dimension of the polytope characterizing the measurements becomes \be D_M = r (n_yn_z- n_y - n_t). \nonumber\ee By contrast, the number of independent variables to characterize the measurements in our method is \be n_yn_z- n_y - n_t \nonumber\ee (see \eqref{m1}-\eqref{m3}). Both the polytopes required in our method, one associated with $\{q(x)\}$ and the one associated with $\{\xi(z|y)\}$, have exponentially smaller dimension in terms of $n_y$ than the polytopes associated with $\{\mu(\lambda|x)\}$ and $\{\xi(z|y,\lambda)\}$. Moreover, to generate $\mathbbm{P}_{\text{P}}$, our method requires only to evaluate the tensor product of the polytope associated with $\{q(x)\}$ and the one associated with $\{\xi(z|y)\}$, eliminating the requirement of projection. This significantly reduces the computational steps. 

\blk 

\section{Noncontextuality inequalities for various scenarios and their quantum violations}

In this section, we explore various elementary contextuality scenarios in detail. First, we retrieve all the noncontextuality inequalities using the aforementioned method. Subsequently, we conduct a comprehensive study of their quantum violations and establish upper bounds on the maximum quantum violations. Before presenting the results, we introduce a \textit{lemma} that greatly aids in obtaining quantum violation for any preparation noncontextuality inequality with binary outcomes.

\begin{lemma}\label{lemma:ms}
Consider any linear figure of merit in any contextuality scenario where there are no nontrivial indistinguishability conditions on measurements, and the outcomes are binary, that is, $z \in \{0,1\}$. Since the outcomes are binary, we can replace $p(1|x,y)$ by $1-p(0|x,y)$ and express any linear expression as $\sum_{x,y} c_{x,y} p(0|x,y)$ where $c_{x,y}$ are some real numbers. Given any quantum preparation strategy $\{\rho_x\}$ satisfying the indistinguishability conditions, the best possible quantum measurement strategy $\{\mathbbm{M}_{z|y}\}$ is fully determined by the preparation strategy as
\be   
\mathbbm{M}_{0|y} = \sum_{a >0} \mathbbm{P}_a ,
\ee   
where $\mathbbm{P}_a$ is the eigenprojector of the operator $\left(\sum_x c_{x,y} \rho_x \right) $ corresponds to positive eigenvalue $a$, that is, $\left(\sum_x c_{x,y} \rho_x \right) \mathbbm{P}_a = a \mathbbm{P}_a $.
\end{lemma}
\begin{proof}
By splitting the sum and replacing the probabilities with quantum states and measurements, we find that 
\begin{align}
\sum_{x,y} c_{x,y} p(0|x,y) & =  \sum_y \left( \sum_x c_{x,y} p(0|x,y) \right) \nonumber \\& =  \sum_y \tr \left[ \left( \sum_x c_{x,y} \rho_x \right) \mathbbm{M}_{0|y} \right] .
\end{align}
Given $\{\rho_x \}$, the quantity $\sum_x c_{x,y} \rho_x $ is fixed and Hermitian, having only real eigenvalues. Thus, for every $y$, the best possible quantum measurement strategy is to take $\mathbbm{M}_{0|y}$ as the sum of eigenprojectors corresponding to the positive eigenvalues of $\sum_x c_{x,y} \rho_x $. Finally, we have $\mathbbm{M}_{1|y} = \I - \mathbbm{M}_{0|y}$.
\end{proof}

To identify quantum violations, two semi-definite programming-based  methods \cite{Chaturvedi2021characterising} are employed. The first method entails alternating sequences of semi-definite programs and is commonly referred to as the see-saw method. Originally introduced in \cite{Pal-Vertesi} for studying the violation of Bell inequalities, this method has subsequently found applications in various other contexts. It provides a lower bound on the maximal quantum violation of the noncontextuality inequalities, along with the corresponding quantum states and measurements that lead to such violation. This optimization involves the following steps. Initially, random quantum states of fixed dimension are generated. Quantum measurements are then optimized to maximize the relevant linear function of probabilities (the left-hand-side of the relevant noncontextuality inequality) while satisfying the indistinguishability conditions on measurements. In general, this set forms a semi-definite program, except in cases where there are no indistinguishability conditions on measurements and outcomes are binary, and \textit{Fact} \ref{lemma:ms} is used to determine the optimal quantum measurements for the generated quantum states.
In the subsequent step, the optimized measurements from the previous step are fixed, and the best quantum states are found that optimize the relevant expression while satisfying indistinguishability conditions on preparations. Yet again, this step forms a semi-definite program. This two-fold optimization process is iterated until the value of the relevant expression saturates. Moreover, this entire optimization is performed for different choices of initial random quantum states, and the best value along with the associated quantum strategy among these is retained. The value obtained through this see-saw method is denoted as $\mathcal{Q}^d_s$ for the chosen dimension ($d$) of the quantum states and measurements. While the value $\mathcal{Q}^d_s$ may not be the optimal quantum value, the method is useful as it delivers the corresponding quantum states and measurements that achieve this value. We have studied the value of $\mathcal{Q}_s^d$ at least upto $d=4$ in all the scenarios. If the obtained value is constant across several dimensions, we enlist the value for minimum possible $d$.

Additionally, the robustness of the quantum violations is studied as a means of comparing violations of different noncontextuality inequalities. Here, we consider the robustness with respect to the presence of white noise, which is the maximum amount of white noise that can be added while the quantum violation persists. Specifically, given the quantum states $\{\rho_x\}_x$ and quantum measurements $\{\mathbbm{M}_{z|y}\}_{z,y}$ acting on $\mathbbm{C}^d$ achieving $\Q^d_s$ found from the see-saw method, we take the noisy states 
$\omega (\I/d) + (1-\omega) \rho_x $, where $\omega \in [0,1]$ being the noise parameter. As the measure of robustness, the minimum value of $\omega$, denoted by $\omega^d_c$, is then determined so that the left-hand side of noncontextuality inequality coincides with the noncontextual bound $\C$. It can be readily verified that, for a general inequality \eqref{gfi},
\be 
\omega^d_c = \frac{\Q^d_s - \C }{\Q^d_s - \gamma} ,
\ee 
where $\gamma = (1/d) \sum_{x,y,z} c_{x,y,z} \tr ( \mathbbm{M}_{z|y} )$ is the value of left-hand-side of \eqref{gfi} for the maximally mixed state. The larger value of $\omega^d_c$ the more robust is the quantum violation arising from $\{\rho_x\}_x$ and $\{\mathbbm{M}_{z|y}\}_{z,y}$.

The second method involves implementing the semi-definite hierarchy introduced in \cite{chaturvedi2021quantum}, which yields a dimension independent upper bound on the maximum quantum violation of noncontextuality inequality. We report this hierarchy up to the first level and denote this upper bound as $\Q_1$. When $\Q_1$ matches with $\Q^d_s$, it signifies the exact maximum violation (up to machine precision). Since, for most inequalities obtained here, the $\Q_1$ values are very close to $\Q^d_s$, higher-order levels were not considered.





The simplest scenario of four preparations satisfying indistinguishability conditions \eqref{ec21porac} has been discussed rigorously in Sec. \ref{sec:method}. Eight other scenarios with their quantum violations are discussed in the following subsections. The respective polytopes and symmetries have been found using \texttt{`polymake'}, and the bounds on the quantum violations based on the aforementioned semi-definite programming methods are found using \texttt{`Matlab'} and \texttt{`sdpt3'}. The codes are available in \cite{codes}.
We enlist the noncontextuality inequalities except the \textit{trivial} ones that are of the form $p(z|x,y) \leqslant 1$ or $p(z|x,y) \geqslant 0 $. Apart from the last scenario, all other scenarios involve only binary outcomes. For the sake of convenience thereof, we express the noncontextuality inequalities only in terms of outcome $z=0$ and further use the following short-hand notation to denote the probabilities
\be 
p_{x,y} := p(0|x,y) .
\ee  

\onecolumngrid

\subsection*{Scenario 2}
The contextuality scenario is defined as follows:
\begin{align}\label{s2}
x \in \{0,1,2,3\},\ y \in \{0,1,2\}, \ z \in \{0,1\}, \quad 
\frac{1}{3} \left( P_0 + P_1 + P_2 \right) \sim \frac{1}{2} \left( P_0 + P_3 \right) . 
\end{align}
The set of inequivalent nontrivial noncontextuality inequalities is enlisted in Table \ref{sc2table}, and the quantum strategy that achieves the highest quantum violation of $\mathcal{I}_2$ is illustrated in Fig. \ref{fig:qv}.
%
\begin{center}
\begin{table}[!ht]
\centering
    \begin{tabular}{c|c|c|c|c}
    \hline 
      orbit size & Inequalities & $\Q_s^{2}$ & $\omega^2_c$ & $\Q_1$ \\
         \hline 
         \hline 
        24 & $\mathcal{I}_2 = - p_{0,0} + 2p_{1,0} + p_{0,1} - 2p_{2,1} \leqslant 2$ & 2.6458 & 0.244 & 2.7321   \\
       \hline
       5 & $ p_{0,2} - 2p_{1,2} - 2p_{2,2} \leqslant 0$ & 0 & 0 & 0  \\
       \hline
    \end{tabular}
    \caption{We obtained $48$ inequalities for the scenario in \eqref{s2}. Out of these inequalities, $19$ are trivial. The rest of them are reduced to $2$ inequivalent classes after applying the indistinguishable conditions and symmetries mentioned below.}
    \label{sc2table}
    \small{  
    $
\forall y,z, \ p(z|x=3,y) =  \frac{2}{3} p(z|x=1,y) + \frac23  p(z|x=2,y) - \frac13 p(z|x=0,y) ; $\\$ P_1 \longleftrightarrow P_2 ; 
   \ M_{0|y} \longleftrightarrow M_{1|y} \ \forall y ; \ \  M_{z|0} \longleftrightarrow M_{z|1}; \ \ M_{z|1} \longleftrightarrow M_{z|2};
    \ M_{z|0} \longleftrightarrow M_{z|2} \forall z . 
    $ }
\end{table}        
\end{center}
%
An interesting aspect of this contextuality scenario deserves attention. We define an ontic distribution as deterministic when $\mu(\lambda|x) \in \{0,1\}$; otherwise, it's probabilistic. Notably, the bound 2 of the noncontextuality inequality $\mathcal{I}_2$ (in Tab.\ref{sc2table}) can only be attained through a probabilistic ontic distribution in any ontological model. For instance, consider the following model where $  \lambda_1,\lambda_2 \in \Lambda $:
\begin{table}[!ht]
    \centering
    \begin{tabular}{c|c c c c}
    \hline 
      $x$ & 0 & 1 & 2 & 3 \\
         \hline 
         $\mu(\lambda_1|x)$ & 0 & 1 & 0 & $\frac23$ \\
         \hline
         $\mu(\lambda_2|x)$ & 1 & 0 & 1 & $\frac13$ \\
       \hline 
    \end{tabular} \ \ \ 
    \begin{tabular}{c| c}
    \hline
      $p_d(0|0,\lambda_1)$ & 1 \\
      \hline
      $p_d(0|0,\lambda_2)$ & 0 \\
      \hline 
      $p_d(0|1,\lambda_2)$ & 0 \\
         \hline 
    \end{tabular}
\end{table} 

This model satisfies \be \label{scen2m}
\frac13 (\mu(\lambda|0)+\mu(\lambda|1)+\mu(\lambda|2)) = \frac12 (\mu(\lambda|0)+\mu(\lambda|3)) .
\ee while simultaneously yielding $\mathcal{I}_2 = 2$. On the other hand, to determine the maximum value of $\mathcal{I}_2$ for deterministic epistemic states, it is sufficient to examine all feasible deterministic epistemic states considering at most four distinct ontic states. The optimization reveals that the value of $\mathcal{I}_2$ is $1$. This observation offers an interesting insight that any value of $\mathcal{I}_2$ greater than 1 indicates the presence of epistemic randomness within the noncontextual ontological model.


\subsection*{Scenario 3}
 The contextuality scenario consists of five preparations and two binary outcome measurements and is defined as follows:
\be \label{s3}
x \in \{0,1,2,3,4\},\ y \in \{0,1\}, \ z \in \{0,1\}, \quad 
\frac{1}{2} \left( P_0 + P_1 \right) \sim \frac{1}{3} \left( P_2 + P_3 + P_4 \right). 
\ee 
The set of inequivalent noncontextuality inequalities for the contextuality scenario is given in Table \ref{sc3table}, and the strategy that attains maximum quantum violation of $\mathcal{I}_3$ is illustrated in Fig. \ref{fig:qv}. \newpage
   
\begin{center}
\begin{table}[!ht]
    \centering
    \begin{tabular}{c|c|c|c|c}
    \hline 
      orbit size & Inequalities & $Q_s^{2}$ & $\omega^2_c$ & $Q_1$ \\
         \hline 
         \hline 
         2 & $-3p_{0,0} - 3p_{1,0} + 2 p_{2,0} + 2p_{3,0} \leqslant 0$ & 0 & 0 & $0$ \\ 
       \hline 
       24 & $\mathcal{I}_3 = - 3p_{0,0} + 2p_{2,0} - 3p_{1,1} + 2p_{2,1} \leqslant 2$ & 3.1231 & 0.272 & $3.1815$\\ 
       \hline 
    \end{tabular}
    \caption{ We obtained $44$ inequalities in this scenario, among which $18$ are trivial. The rest of the inequalities are reduced to $2$ inequivalent classes after employing the following indistinguishability conditions and symmetries.}
    \label{sc3table}
\small{
$
  p(z|x=4,y) =  \frac{3}{2}\left( p(z|x=0,y) + p(z|x=1,y)\right) - p(z|x=2,y) - p(z|x=3,y) \ \forall y,z
$\\$
    P_0 \longleftrightarrow P_1 ; 
    P_2 \longleftrightarrow P_3;
    P_2 \longleftrightarrow P_4;
    P_3 \longleftrightarrow P_4 $;\\$
    \ M_{0|y} \longleftrightarrow M_{1|y} \ \forall y ;   M_{z|0} \longleftrightarrow M_{z|1} \ \forall z ;
$\\$
\label{sc3m}
    \ M_{0|0} \longrightarrow M_{0|1} \ , \ M_{0|1} \longrightarrow M_{1|0} \ ,\ M_{1|0} \longrightarrow M_{1|1} \ ,\ M_{1|1} \longrightarrow M_{0|0} .
$\\
The symmetry operation $M_{z|y}\longrightarrow M_{z'|y'}$ corresponds to the relabeling of the variables $z,y$ to $z',y'$ that together $\forall y,z$ describes a symmetry of this scenario. In what follows, symmetry operations separated by $\{,\}$ are meant to be applied together.}  
\end{table}
\end{center}  \vspace{-1.5 cm}
\subsection*{Scenario 4}
Table \ref{sc4table} contains the inequivalent noncontextuality inequalities for the contextuality scenario defined as follows:
\be
x \in \{0,1,2,3,4\}, y \in \{0,1\},z \in \{0,1\} \quad
 \frac{1}{4} \left( P_0 + P_1 + P_2 + P_3 \right) \sim \frac{1}{3} \left( P_0 + P_1 + P_4 \right) .
\ee

\begin{center}
\begin{table}[!ht]
    \centering
    \begin{tabular}{c|c|c|c|c}
    \hline 
      orbit size & Inequalities & $Q_s^{2}$ & $\omega^2_c$ & $Q_1$ \\
         \hline 
         \hline 
         4 & $ p_{0,0} + p_{1,0} - 3 p_{2,0} - 3p_{3,0} \leqslant 0$ & 0 & 0 & 0 \\
         \hline
         8 & $ p_{0,0} + p_{1,0} - 3 p_{2,0} + p_{0,1} + p_{1,1} - 3p_{3,1} \leqslant 2 $ & 3.1231 & 0.272 & $3.1815$ \\
         \hline
         16 & $ p_{1,0} -3 p_{2,0} + p_{1,1} - 3p_{3,1} \leqslant 1$ & 1.7417 & 0.198 & $1.9168$ \\
         \hline
    \end{tabular}   
    \caption{We obtained $44$ inequalities including 16 trivial ones. The 28 nontrivial noncontextuality inequalities are reduced to 3 inequivalent classes after applying the following indistinguishability conditions and symmetries.}
    \label{sc4table}
\small{$ 
 \ p(z|x=4,y) =  \frac{3}{4} ( p(z|x=2,y) + p(z|x=3,y) ) -\frac{1}{4} ( p(z|x=0,y) + p(z|x=1,y) ) \ \forall y,z
$\\$    
    P_0 \longleftrightarrow P_1 ; 
    P_2 \longleftrightarrow P_3 ; $\\$
    \ M_{0|y} \longleftrightarrow M_{1|y} \ \forall y ; \ \  M_{z|0} \longleftrightarrow M_{z|1} \ \forall z;
$\\$
\label{sc4}
    \ M_{0|0} \longrightarrow M_{0|1} \ , \ M_{0|1} \longrightarrow M_{1|0} \ ,\ M_{1|0} \longrightarrow M_{1|1} \ ,\ M_{1|1} \longrightarrow M_{0|0} .
$
}
\end{table}    
\end{center}\vspace{-1.2 cm}
\subsection*{Scenario 5}
The contextuality scenario consisting of six preparations and three binary outcome measurements is defined as:
\be
x \in \{0,1,2,3,4,5\}, y \in \{0,1,2\},z \in \{0,1\},  \quad \frac{1}{2} \left( P_0 + P_1 \right) \sim \frac{1}{2} \left( P_2 + P_3 \right) \sim \frac{1}{2} \left( P_4 + P_5 \right) .
\ee 
The set of inequivalent noncontextuality inequalities is given in Table \ref{sc5table}, and the qubit strategy that achieves the maximum quantum violation of $\mathcal{I}_5$ is provided in Fig. \ref{fig:qv}. 
\begin{center}
\begin{table}[!htbp]
    \centering
    \begin{tabular}{c|c|c|c|c}
    \hline 
      orbit size & Inequalities & $Q_s^{2}$ & $\omega^2_c$ & $Q_1$ \\
         \hline 
         \hline 
         24 & $ p_{0,0} + p_{1,0} - p_{4,0} \leqslant 1$ & 1 & 0 & 1\\
         \hline
         24 & $ p_{0,0} - p_{2,0} - p_{1,2} + p_{2,2} \leqslant 1$ & 1.4142 & 0.293 & 1.4142 \\
         \hline
         24 & $ p_{1,0} - p_{4,0} + p_{0,2} - p_{4,2} \leqslant 1$ & 1.4142 & 0.293 & 1.4142\\
         \hline
         24 & $ p_{2,1} - p_{4,1} - p_{0,2} - p_{1,2} + p_{2,2} + p_{4,2} \leqslant 1$ & 1.4142 & 0.293 & 1.4142\\
         \hline
        192 & $ \mathcal{I}_5 = p_{2,0} - p_{4,0} + p_{1,1} - p_{2,1} - p_{4,1} - p_{0,2} + p_{2,2} + p_{4,2} \leqslant 2$ & 2.5 & 0.2 & 2.6571\\
        \hline
          192 & $ - p_{0,0} + p_{2,0} + p_{4,0} + p_{1,1} - p_{2,1} - p_{1,2} - p_{2,2} + p_{4,2} \leqslant 2$ & 2.5  & 0.2 & 2.6571\\
         \hline
          192 & $ p_{1,0} - p_{4,0} + p_{0,1} + p_{2,1} - p_{4,1} - 2p_{0,2} - p_{1,2} + p_{2,2} + p_{4,2}\leqslant 2$ & 2.5  & 0.2 & 2.6571\\
         \hline
    \end{tabular}   
    \caption{Here we obtained 684 inequalities. Among these, 12 are trivial. The remaining inequalities are reduced to \re $7$ \blk  inequivalent classes after applying the following indistinguishability conditions and symmetry transformations.}
    \label{sc5table}
\small{
$ 
 \ p(z|x=3,y) =  p(z|x=0,y) + p(z|x=1,y) - p(z|x=2,y) \ \forall y,z 
$\\$ 
 \ p(z|x=5,y) =  p(z|x=0,y) + p(z|x=1,y) - p(z|x=4,y) \ \forall y,z 
$\\$
    P_0 \longleftrightarrow P_1 ; 
    P_2 \longleftrightarrow P_3 ;
    P_4 \longleftrightarrow P_5 ;
$ \vspace{-0.3cm}
\begin{align}
\label{symM5}
   \ M_{0|y} \longleftrightarrow M_{1|y}\ \forall y ;
    \ M_{z|0} \longleftrightarrow M_{z|1};
    \ M_{z|1} \longleftrightarrow M_{z|2};
    \ M_{z|0} \longleftrightarrow M_{z|2}
   \ \forall z .
\end{align}
}
\end{table}
\end{center} 
\vspace{-2 cm}

\subsection*{Scenario 6}
The contextuality scenario consisting of seven preparations and three measurements is defined as follows:
\be
x \in \{0,1,2,3,4,5,6\}, y \in \{0,1,2\},z \in \{0,1\} , \quad  
 \frac{1}{4} \left( P_0 + P_1 + P_2 + P_3 \right) \sim \frac{1}{3} \left( P_4 + P_5 + P_6 \right) .
\ee 
Table \ref{sc6table} presents the collection of inequivalent noncontextuality inequalities obtained in this scenario. In contrast to previous scenarios, qubit strategy in this context did not yield the optimal results (using the see-saw method). In particular, there is a substantial enhancement in $Q_s$ when considering states and measurements of higher dimensions. We have documented these values up to $d=4$. Notably, for $\mathcal{I}_6^2$ and $\mathcal{I}_6^4$, we obtain $Q^7_s=14.9711$ and $Q_s^7 = 15.6163$, respectively, which are more than $Q_s^4$ given in table \ref{sc6table}. In Fig. \ref{fig:qv}, we provide the details of the qubit strategy which violates $\mathcal{I}_6^3$. Here, we present the $3$-dimensional quantum states and measurements that yield a quantum violation of $\mathcal{I}_6^1 = 8.7764>6$:
\begin{gather}
\rho_0 = 
\Bigg(\begin{smallmatrix}
    0.7865 &  -0.1091  &  0.3950 \\
   -0.1091  &  0.0151  & -0.0548  \\
    0.3950 &  -0.0548  &  0.1983 
\end{smallmatrix}\Bigg),
\rho_1=\Bigg(\begin{smallmatrix}
    0.0004 &   0.0186  &  0.0077 \\
    0.0186 &   0.8526  &  0.3540 \\
    0.0077 &   0.3540  &  0.1470
\end{smallmatrix}\Bigg),
\rho_2 = \Bigg(\begin{smallmatrix}
    0.0004  &  0.0186  &  0.0077 \\
    0.0186  &  0.8526  &  0.3540 \\
    0.0077  &  0.3540  &  0.1470
\end{smallmatrix}\Bigg), \nonumber \\
\rho_3= \Bigg(\begin{smallmatrix}
    0.3904  &  0.2225 &  -0.4341 \\
    0.2225  &  0.1268 &  -0.2474 \\
   -0.4341  & -0.2474 &   0.4828
\end{smallmatrix}\Bigg), \rho_4=\Bigg(\begin{smallmatrix}
    0.1713  &  0.2578 &   0.1378 \\
    0.2578  &  0.5426 &   0.0428 \\
    0.1378  &  0.0428 &   0.2860
\end{smallmatrix}\Bigg),
\rho_5=\Bigg(\begin{smallmatrix}
     0.5406 &  -0.4028  & -0.2935 \\
   -0.4028  &  0.3000   & 0.2187 \\
   -0.2935  &  0.2187   & 0.1593
\end{smallmatrix}\bigg), \nonumber \\ \rho_6= \Bigg(\begin{smallmatrix}
       0.1713  &  0.2579  &  0.1378 \\
    0.2579  &  0.5427  &  0.0429 \\
    0.1378 &   0.0429  &  0.2860
\end{smallmatrix}\Bigg) ,\nonumber \\ 
\mathbbm{M}_{0|0}= \Bigg(\begin{smallmatrix}
     0.8633 &  -0.3370 &   0.0668 \\
   -0.3370  &  0.1315  & -0.0261 \\
    0.0668 &  -0.0261  &  0.0052
\end{smallmatrix}\Bigg), \
\mathbbm{M}_{0|1}=\Bigg(\begin{smallmatrix}
    0.5859 &  -0.0485  & -0.4902 \\
   -0.0485 &   0.0040  &  0.0406 \\
   -0.4902 &   0.0406  &  0.4101
\end{smallmatrix}\Bigg), \
\mathbbm{M}_{0|2}=\Bigg(\begin{smallmatrix}
   0.7889  &  0.3566  &  0.1984 \\
    0.3566 &   0.3976 &  -0.3351 \\
    0.1984 &  -0.3351 &   0.8136
\end{smallmatrix}\Bigg) .
\end{gather}
Moreover, the following set of $4-$dimensional quantum states and measurements achieve the quantum violation $\Q_s = 15.5037>12$ of $\mathcal{I}_6^4$ in Tab. \ref{sc6table},
\begin{gather}
\rho_0 = 
\Bigg(\begin{smallmatrix}
    0.1537 &  -0.1618 &  -0.1204 &  -0.2990 \\
   -0.1618 &   0.1704 &   0.1268 &   0.3148 \\
   -0.1204 &   0.1268 &   0.0943 &   0.2342 \\
   -0.2990 &   0.3148 &   0.2342 &   0.5816
\end{smallmatrix}\Bigg),
\rho_1=\Bigg(\begin{smallmatrix}
    0.3830  & -0.3343 &   0.0919  &  0.3408 \\
   -0.3343  &  0.2917 &  -0.0802  & -0.2974 \\
    0.0919  & -0.0802 &   0.0220  &  0.0818 \\
    0.3408  & -0.2974 &   0.0818  &  0.3032
\end{smallmatrix}\Bigg), \nonumber \\
\rho_2 = \Bigg(\begin{smallmatrix}
    0.1411 &   0.2127 &  -0.1319 &   0.0790 \\
    0.2127 &   0.4379 &  -0.0719  &  0.0026 \\
   -0.1319 &  -0.0719 &   0.2609  & -0.2001 \\
    0.0790 &   0.0026 &  -0.2001  &  0.1601 \\
\end{smallmatrix}\Bigg), \rho_3= \Bigg(\begin{smallmatrix}
    0.1411 &   0.2127 &  -0.1319 &   0.0790 \\
    0.2127 &   0.4379 &  -0.0718 &   0.0026 \\
   -0.1319 &  -0.0718 &   0.2609 &  -0.2001 \\
    0.0790 &   0.0026 &  -0.2001 &   0.1601
\end{smallmatrix}\Bigg), \\
\rho_4=\Bigg(\begin{smallmatrix}
    0.1570  & -0.2543 &  -0.0758 &   0.2489 \\
   -0.2543  &  0.4119 &   0.1227 &  -0.4031 \\
   -0.0758  &  0.1227 &   0.0366 &  -0.1201 \\
    0.2489  & -0.4031 &  -0.1201 &   0.3945
\end{smallmatrix}\Bigg),
\rho_5=\Bigg(\begin{smallmatrix}
    0.0102  &  0.0748 &  -0.0141 &   0.0657 \\
    0.0748  &  0.5477 &  -0.1035 &   0.4811 \\
   -0.0141  & -0.1035 &   0.0196 &  -0.0909 \\
    0.0657  &  0.4811 &  -0.0909 &   0.4225
\end{smallmatrix}\Bigg),\nonumber  \\
\rho_6= \Bigg(\begin{smallmatrix}
     0.4470 &   0.1266  & -0.1294 &  -0.1646 \\
    0.1266  &  0.0438  & -0.0921  & -0.0610 \\
   -0.1294  & -0.0921  &  0.4225  &  0.1478 \\
   -0.1646  & -0.0610  &  0.1478  &  0.0867
\end{smallmatrix}\Bigg), \ 
\mathbbm{M}_{0|0}= \Bigg(\begin{smallmatrix}
    0.0406 &  -0.1190  & -0.0263 &  -0.1552 \\
   -0.1190 &   0.3488  &  0.0772 &   0.4550 \\
   -0.0263 &   0.0772  &  0.0171 &   0.1007 \\
   -0.1552 &   0.4550  &  0.1007 &   0.5935
\end{smallmatrix}\Bigg), \nonumber \\
\mathbbm{M}_{0|1}=\Bigg(\begin{smallmatrix}
    0.1083 &   0.1665 &  -0.1530  &  0.2133 \\
    0.1665 &   0.9667 &   0.0656  & -0.0115 \\
   -0.1530 &   0.0656 &   0.3433  & -0.4447 \\
    0.2133 &  -0.0115 &  -0.4447  &  0.5816
\end{smallmatrix}\Bigg), \
\mathbbm{M}_{0|2}=\Bigg(\begin{smallmatrix}
   0.3099  & -0.3260  &  0.0207  &  0.3274 \\
   -0.3260 &   0.3429 &  -0.0217 &  -0.3444 \\
    0.0207 &  -0.0217 &   0.0014 &   0.0218 \\
    0.3274 &  -0.3444 &   0.0218 &   0.3459
\end{smallmatrix}\Bigg) .
\end{gather}
\newpage

\begin{center}
\begin{table}[!ht]
    \centering
    \small
    \begin{tabular}{c|c|c|c|c|c|c|c|c|c}
    \hline 
      orbit & Inequalities  & $Q_s^{2}$ & $\omega_c^{2} $ & $Q_s^{3}$ & $\omega_c^{3} $ & $Q_s^{4}$ & $\omega_c^{4} $ & $Q_1$ & $Q^2_{UB}$\\
      size & & & & & & & & \\
         \hline 
         \hline 
         4 & $ 3p_{0,0} + 3p_{1,0} + 3p_{2,0} + 3p_{3,0} $ & 4 & 0 & 4 & 0 & 4 & 0 & 4 & 4\\ 
          & $- 4p_{4,0}- 4p_{5,0} \leqslant 4$ & & & & & & & &

         \\ \hline
         864 & $\mathcal{I}_6^1= - 3p_{1,0} - 3p_{2,0} - 3p_{3,0} $ &  & & & & & & \\ & $ + 4p_{5,0} - 3p_{0,1} - 3p_{1,1}- 3p_{2,1} $ & 7.6833 & 0.144 & 8.7764 & 0.257 & 8.7764 & 0.270 & 9.2621 & 7.6833\\
         & $ + 4p_{5,1} + 3p_{0,2}+ 3p_{3,2} $ & & & & & & & &
         \\ & $- 4p_{5,2}\leqslant 6$ & & & & & & & & \\ \hline
   864 & $\mathcal{I}_6^2 = - 6p_{0,0} - 3p_{2,0} - 3p_{3,0} $ &  & & & & & & \\ & $ + 4p_{4,0} + 8p_{5,0} - 3p_{0,1} - 3p_{1,1} $ & 12 & 0 & 14.4414 & 0.161 & 14.6929 & 0.171 & 16.9615 & 12\\
   & $+ 4p_{4,1}- 6p_{1,2} - 3p_{2,2}- 3p_{3,2} $ &  &  &  & & &  &  \\ & $+ 4p_{4,2} + 8p_{5,2} \leqslant 12$ & & & & & & & & \\
         \hline
         864 & $ - 3p_{3,0} + 4p_{4,0} - 3p_{0,1} - 3p_{2,1}$ & & & & & & & \\ & $ + 4p_{4,1}+ 3p_{1,2} - 4p_{4,2} \leqslant 7$ & 8.9692 & 0.197 & 8.9692 & 0.197 & 9.472 & 0.236 & 10.5888 & 8.9692\\
         \hline
         288 & $ - 3p_{1,0} + 4p_{5,0} + 3p_{0,2} + 3p_{2,2} $ & & & & & & & &\\
         & $+ 3p_{3,2}- 4p_{5,2} \leqslant 9$ & 10.8281 & 0.233 & 10.8281 & 0.267 & 10.8281 & 0.277& 11.1229 & 10.8281 
         \\ \hline
        216 & $ \mathcal{I}_6^3= - 3p_{1,0} - 3p_{3,0} + 4p_{5,0} $ & & & & & & & \\
        & $+ 3p_{0,2}+ 3p_{2,2} - 4p_{5,2} \leqslant 6$ & 8.2462 & 0.272 & 8.2462 & 0.296 & 8.2462 & 0.310 & 8.3631 & 8.2462\\
         \hline
         288 & $ - 3p_{0,0} - 3p_{3,0} + 4p_{4,0} + 4p_{5,0} $ & & & & & & & \\ & $ - 3p_{1,1} - 3p_{2,1}+ 4p_{5,1} - 3p_{1,2}$ & 9.8167 & 0.167 & 11 & 0.272 & 11 & 0.272 & 11.7279 & 9.8167 \\ & $- 3p_{2,2} + 4p_{4,2} \leqslant 8$ & & & & & & & & \\
         \hline
         192 & $ 3p_{1,0} - 4p_{5,0} - 3p_{0,1} - 3p_{2,1} $ & & & & & & & \\ & $ - 3p_{3,1} + 4p_{4,1} + 4p_{5,1} - 3p_{1,2} $ & 8.6241 & 0.178 & 8.6241 & 0.188 & 9.1366 & 0.221 & 10 & 8.6241\\ & $+ 4p_{4,2} \leqslant 7$ & & & & & & & & \\
         \hline
         192 & $ 3p_{1,0} + 3p_{2,0} + 3p_{3,0} - 4p_{4,0} $ & & & & & & & \\ & $- 3p_{1,1} - 3p_{2,1} - 3p_{3,1} + 4p_{5,1} $ & 14.2154 & 0.103 & 15.658 & 0.249 & 15.658 &  0.249 & 15.8923 & 14.2154\\ & $- 3p_{0,2} + 4p_{4,2} + 4p_{5,2} \leqslant 13$ & & & & & & & & \\
         \hline
       1728 & $\mathcal{I}_6^4= - 3p_{1,0} - 6p_{2,0} - 6p_{3,0} $ &  & & & & & & \\ & $ + 4p_{4,0} + 8p_{5,0}- 6p_{0,1} - 3p_{1,1} $ & 12.3578 & 0.024 & 15.1551 & 0.199 & 15.5037 & 0.218 & 17.2706 & 12.3578\\
       & $+ 4p_{4,1} + 8p_{5,1} - 3p_{0,2} - 3p_{2,2}$ &  & & & & & & \\ & $- 3p_{3,2} + 4p_{4,2} \leqslant 12$ & & & & & & & & \\
         \hline
    \end{tabular} 
    \caption{In this case, we obtained $5538$ inequalities, among which $38$ are trivial. The rest of the $5500$ inequalities reduce to $10$ inequivalent class as mentioned above. To obtain the inequivalent noncontextuality inequalities, we apply the following indistinguishability conditions and symmetry transformations.}
    \label{sc6table}
\small{
$ 
  \ p(z|x=6,y) =  \frac{3}{4} \left( p(z|x=0,y) + p(z|x=1,y) + p(z|x=2,y) + p(z|x=3,y) \right) - p(z|x=4,y) - p(z|x=5,y) \ \forall y,z
$\\$
    P_0 \longleftrightarrow P_1; 
    P_0 \longleftrightarrow P_2 ;
    P_0 \longleftrightarrow P_3 ;
    P_1 \longleftrightarrow P_2 ;
    P_1 \longleftrightarrow P_3 ;
    P_2 \longleftrightarrow P_3;
    P_4 \longleftrightarrow P_5;
    P_4 \longleftrightarrow P_6;
    P_5 \longleftrightarrow P_6
;$\\$  
    \ M_{0|y} \longleftrightarrow M_{1|y}
   \ \forall y ;$ \\$ 
    \ M_{z|0} \longleftrightarrow M_{z|1};
    \ M_{z|1} \longleftrightarrow M_{z|2};
    \ M_{z|0} \longleftrightarrow M_{z|2}
   \ \forall z .
$ \\ Here $Q^2_{UB}$ denotes the quantum upper bound for two dimensional systems. 
}
\end{table}
\end{center}

\subsection*{Scenario 7}
This contextuality scenario involves eight preparations and three measurements, that is, $x \in \{0,1,2,3,4,5,6,7\}, y \in \{0,1,2\},z \in \{0,1\}$, satisfying the following indistinguishability conditions:
\be
\begin{aligned}\label{sc7a}
 \frac{1}{4} \left( P_0 + P_1 + P_6 + P_7 \right) \sim \frac{1}{4} \left( P_2 + P_3 + P_4 + P_5 \right)  \sim \frac{1}{4} \left( P_0 + P_2 + P_5 + P_7\right) \sim \frac{1}{4} \left( P_1 + P_3 + P_4 + P_6\right)\\ 
 \sim \frac{1}{4} \left( P_0 + P_3 + P_4 + P_7\right) \sim \frac{1}{4} \left( P_1 + P_2 + P_5 + P_6\right) \sim \frac{1}{4} \left( P_0 + P_3 + P_5 + P_6\right) \sim \frac{1}{4} \left( P_1 + P_2 + P_4 + P_7\right). 
\end{aligned}
\ee 
It is worth noting that the indistinguishability conditions in this scenario are not entirely independent. Interestingly, this scenario resembles the $3$-bit parity oblivious multiplexing task, which has been previously explored in \cite{Spekkens2009}. In this communication task, the sender possesses a $3$-bit string $x = x_{0}x_{1}x_{2}$ selected uniformly, and the receiver's goal is to guess the $y$th bit of ${x}$ while adhering to the constraint that all potential parities of the input bits must remain oblivious in the communication. We  find four inequivalent noncontextuality inequalities in this scenario, as detailed in Tab. \ref{sc7table}. To achieve quantum violations of $\mathcal{I}_7$, a specific strategy can be employed wherein $\rho_x$ corresponds to qubit states representing the vertices of the cube on the Bloch sphere shown in Tab. \ref{sc7table} within the Bloch sphere, and the reciever's measurements are $\sigma_y,\sigma_x,\sigma_z$ for $y=0,1,2$, respectively.

\begin{center}
\begin{table}[!ht]
    \centering
    \begin{tabular}{c|c|c|c|c}
    \hline 
      orbit size & Inequalities & $Q_s^{2}$ & $\omega^2_c$ & $Q_1$ \\
         \hline 
         \hline 
         48 & $ -2p_{0,0} + p_{1,0} + p_{2,0} + p_{4,0} \leqslant 1$ & 1 & 0 & 1\\
         \hline
        144 & $ -p_{0,0} + p_{2,0} - p_{0,2} + p_{1,2} \leqslant 1$ & 1.4142  & 0.293 & 1.4142\\
         \hline
        144 & $ 2p_{0,0} - p_{2,0} - 2p_{4,0} - p_{0,1} + p_{1,1} \leqslant 1$ & 1.3371  & 0.183 & 1.3638\\
         \hline
     48 & $ \mathcal{I}_7 = p_{0,0} - p_{1,0} + p_{0,1} - p_{4,1} + p_{0,2} - p_{2,2}\leqslant 1$ & 1.7321 & 0.423 & 1.7321\\
         \hline
    \end{tabular}
    \caption{We obtained $384$ nontrivial inequalities that are grouped into $4$ inequivalent classes. It turns out that by rearranging the indistinguishability conditions, we can express the probabilities for four input variables using the other four input variables in the following manner.}
\label{sc7table}
\vspace{-0.5cm}
\small{\begin{gather}  \label{redp3}
   p(z|x=3,y) =  p(z|x=1,y) + p(z|x=2,y) - p(z|x=0,y) \ \forall y,z  
   \\ 
        p(z|x=5,y) =  p(z|x=1,y) + p(z|x=4,y) - p(z|x=0,y) \ \forall y,z
\\ 
    p(z|x=6,y) =  p(z|x=2,y) + p(z|x=4,y) - p(z|x=0,y) \ \forall y,z \\
 \label{redp7}
   p(z|x=7,y) =  p(z|x=1,y) + p(z|x=2,y) + p(z|x=4,y) - 2p(z|x=0,y) \ \forall y,z.
\end{gather} \\
Interestingly, the symmetries possessed by the preparations $\{P_x\}$ exhibit a fascinating connection to the symmetries of a cube. A cube contains $23$ symmetry elements that include centre of symmetry, plane of symmetry and axis of symmetry. The symmetry operations on the preparations along with the corresponding cube symmetries that we apply to find the equivalent noncontextuality inequalities are described below.  
\begin{center}
\vspace{0.2cm}
    \includegraphics[scale=0.45]{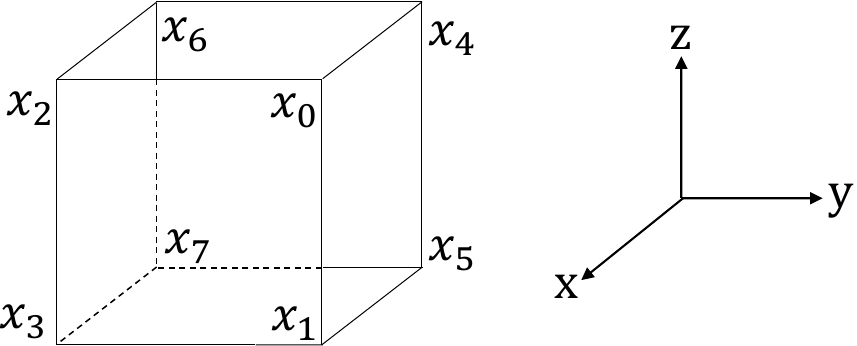}
\end{center}
$
  P_0 \longleftrightarrow P_1, 
    P_6 \longleftrightarrow P_7 ,
    P_2 \longleftrightarrow P_3 ,
    P_4 \longleftrightarrow P_5 ; \text{(reflection with respect to x-y plane)} $ \\$
    P_0 \longleftrightarrow P_4, 
    P_3 \longleftrightarrow P_7 ,
    P_2 \longleftrightarrow P_6, 
    P_1 \longleftrightarrow P_5 ;
 \text{(reflection with respect to y-z plane)} $\\$
    P_0 \longleftrightarrow P_7, 
    P_3 \longleftrightarrow P_4 ,
    P_2 \longleftrightarrow P_6, 
    P_1 \longleftrightarrow P_5 ;
 \text{($\pi$ rotation about an axis dissecting the edges containing $x_2x_6$ and $x_1x_5$)}  $\\$
    P_0 \longleftrightarrow P_4, 
    P_3 \longleftrightarrow P_7 ,
    P_2 \longleftrightarrow P_5, 
    P_1 \longleftrightarrow P_6 ;  
    \text{($\pi$ rotation about an axis dissecting the edges containing $x_3x_7$ and $x_0x_4$)}  $\\$
    P_0 \longleftrightarrow P_7, 
    P_3 \longleftrightarrow P_4 ,
    P_2 \longleftrightarrow P_5, 
    P_1 \longleftrightarrow P_6 ; \text{(inversion between opposite points about centre of symmetry)}$\\$
    P_0 \longleftrightarrow P_6, 
    P_1 \longleftrightarrow P_7 ;
 \text{(reflection with respect to the diagonal plane containing $x_2,x_3,x_5,x_4$)}
 $ \\
Note that there are more symmetries, but the ones listed above are enough to group all the NCI into the inequivalent classes. The symmetry transformations implemented on the measurements are given by \eqref{symM5}.
}
\end{table}
\end{center}

\subsection*{Scenario 8} 
Until this point, we have considered contextuality scenarios involving indistinguishability conditions on preparations. However, in this particular scenario, we examine eight preparations, with $x \in \{0,1,2,3,4,5,6,7\}$ adhering to the same conditions as specified in \eqref{sc7a}, and three binary outcome measurements, $y \in \{0,1,2\}, z \in \{0,1\}$, which satisfy the indistinguishability condition,
\begin{equation}
    \frac{1}{3} \{ M_{0|0}+M_{0|1}+M_{0|2} \} \sim \frac{1}{3} \{ M_{1|0}+M_{1|1}+M_{1|2} \} .
\end{equation}
The resulting inequivalent noncontextuality inequalities can be found in Tab. \ref{sc8table}. An important observation here is that, for all the noncontextuality inequalities, $Q^2_s = Q_1$ up to machine precision, indicating we have found the exact maximum quantum violations. Moreover, $Q_1^\Pi = Q_1$, implying that projective measurements are sufficient to obtain the maximum quantum violations of all the noncontextuality inequalities. A set of qubit states and measurements yielding the maximal violation of $\mathcal{I}_8$ is provided in Fig. \ref{fig:qv}.
\newpage

\begin{table}[!ht]
    \centering
    \begin{tabular}{c|c|c|c|c}
    \hline 
      orbit size & Inequalities & $Q_s^{2}$ & $\omega^2_c$ & $Q_1$  \\
         \hline 
         \hline 
         48 & $-4p_{0,0} + 2p_{1,0} + 2p_{2,0} + 2p_{4,0} -4 p_{0,1}+2 p_{1,1} +2 p_{2,1}+2 p_{4,1}\leqslant 3$ & 3 & 0 & 3 \\ 
        \hline 
         288 & $ -2p_{0,0} +2 p_{1,0} +2 p_{2,0} - p_{0,1} + 2p_{1,1}\leqslant 3$ & 3.3660 & 0.196 & 3.3660  \\ 
       \hline
       288 & $ 5p_{0,0} -3 p_{1,0} -2 p_{2,0} -2 p_{4,0} +2 p_{0,1} -2p_{4,1} \leqslant 1$ & 1.6458 & 0.244 & 1.6458  \\ 
       \hline
       288 & $ p_{1,0} - p_{2,0} + p_{4,0} + 2p_{0,1} \leqslant 3$ & 3.3660 & 0.196 & 3.3660 \\ 
       \hline
        144 & $ -p_{0,0} + p_{2,0} - p_{0,1} + p_{4,1} \leqslant 1$ & 1.3660 & 0.268 & 1.3660\\ 
       \hline
       144 & $ -4p_{0,0} + p_{1,0} + p_{4,0}-2 p_{0,1} +2 p_{2,1} \leqslant 1$ & 1.6458 & 0.244 & 1.6458  \\ 
       \hline
       144 & $ p_{0,0} -3 p_{2,0} -4 p_{0,1}+2 p_{1,1} +2 p_{4,1} \leqslant 1$ & 1.6458 & 0.244 & 1.6458 \\ 
       \hline
       144 & $ -3p_{0,0} + p_{1,0} + p_{2,0}+ p_{4,0} -2 p_{0,1} + p_{1,1} + p_{4,1}\leqslant 1$ & 1.3660 & 0.268 & 1.3660 \\ 
       \hline
     48 & $ \mathcal{I}_8 = -p_{1,0} + p_{4,0} -2 p_{0,1}+ p_{2,1}+ p_{4,1} \leqslant 1$ & 1.5 & 0.333 & 1.5  \\ \hline
       576 & $ p_{0,0} -2 p_{1,0}+ p_{2,0} -4 p_{0,1}+3 p_{2,1}+3 p_{4,1} \leqslant 3$ & 3.6889 & 0.256 & 3.6889  \\ \hline
       576 & $ 4p_{0,0} -4 p_{1,0}-2 p_{4,0} -5 p_{0,1}+4 p_{2,1}+3 p_{4,1} \leqslant 3$ & 3.9210 & 0.235 & 3.9210 \\ \hline
       
    \end{tabular}
    \caption{We obtained $2688$ nontrivial noncontextuality inequalities that reduced to $11$ inequivalent classes. For the preparations, we apply the same indistinguishability relations given by \eqref{redp3}-\eqref{redp7} as in Scenario 7. For the measurements, the following relation is imposed:
    \label{sc8table}
    \small $ 
  p(z=0|x,y=2) = 3/2 - p(z=0|x,y=1) - p(z=0|x,y=0) \ \forall x . $ Moreover, we apply the same set of symmetry transformations on both preparations and measurements, as in Scenario 7.} 

\end{table} 

\subsection*{Scenario 9}

Finally, we consider a scenario where the measurements have three possible outcomes, $z \in \{0,1,2\}$. The indistinguishability conditions in this scenario, consisting of six preparations and two measurements, are given by,
\be 
 \frac{1}{2} \left( P_0 + P_1 \right) \sim \frac{1}{2} \left( P_2 + P_3 \right) \sim \frac{1}{2} \left( P_4 + P_5 \right) ; \quad  
\frac{1}{2} \{ M_{0|0}+M_{0|1} \} \sim \frac{1}{2} \{ M_{1|0}+M_{1|1} \} ,
\ee
where $x \in \{0,1,2,3,4,5\}, y \in \{0,1\}$. 
The set of inequivalent noncontextuality inequalities is given in Tab. \ref{sc9table}, and the quantum violation of $\mathcal{I}_3$ for qubit systems is illustrated in Fig. \ref{fig:qv}. 
\begin{table}[!ht]
\small
    \centering
    \begin{tabular}{c|c|c|c|c|c}
    \hline 

      orbit size & Inequalities & $\Q_s^{2}$ & $\omega^2_c$ & $\Q_1$ & $\Q^\Pi_1$ \\
         \hline 
         \hline 
         6 & $p^{0}_{4,0} +2 p^{0}_{4,1} - p^{1}_{4,0} \leqslant 1$ & 1 & 0 & 2 & 0 \\
         \hline 
         6 &  $-2 p^{0}_{1,0} -2 p^{0}_{1,1} +2 p^{1}_{1,0} \leqslant 0$ & 0 & 0 & 0 & 0 \\
         \hline
         6 & $-2 p^{1}_{0,0} -2 p^{1}_{1,0} +2 p^{1}_{4,0} \leqslant 0$ & 0  & 0 & 0 & 0 \\
         \hline 
         8 & $-2 p^{0}_{0,0} - 2 p^{0}_{1,0} +3 p^{0}_{2,0} -2 p^{0}_{0,1} - 2 p^{0}_{1,1} + 2 p^{0}_{2,1}+$ &  & & \\ & $2 p^{0}_{4,1} + 2 p^{1}_{0,0} +2 p^{1}_{1,0} - p^{1}_{2,0} - 2 p^{1}_{4,0}\leqslant 1$ & 1.4536  & 0.312 & 2.8284 & 1 \\
         \hline
         8 & $-2p^{0}_{0,0} -2 p^{0}_{1,0} +3 p^{0}_{4,0} -2p^{0}_{2,1} + 2p^{0}_{4,1} + 2p^{1}_{2,0} - p^{1}_{4,0}\leqslant  1$ & 1.4536  & 0.312 & 2.8284 & 1 \\
         \hline
        16 & $ 
\mathcal{I}_9 =  p^{0}_{1,0} +2 p^{0}_{1,1}  -2p^{0}_{2,1} -2 p^{1}_{0,0} -p^{1}_{1,0} +2 p^{1}_{2,0} \leqslant 1$ & 1.4536  & 0.312  & 2.8284 & 1 \\
         \hline
         8 & $-2p^{0}_{0,0} -2 p^{0}_{0,1} -2 p^{0}_{1,1} +2p^{0}_{4,1} + 2p^{1}_{0,0} - 2p^{1}_{4,0} \leqslant 0 $ & 0.5  & 0.2 & 0.8284 & 0 \\
         \hline
         16 & $-2p^{0}_{0,0} -2 p^{0}_{1,0} +3 p^{0}_{4,0} -2p^{0}_{1,1} + 2p^{0}_{4,1} + 2p^{1}_{1,0} - p^{1}_{4,0} \leqslant 1$ & 1.4536  &  0.312 & 2.8284 & 1 \\ 
         \hline
      8 & $ 
-2p^{0}_{0,0} -2 p^{0}_{1,0} +2 p^{0}_{4,0} -2p^{0}_{1,1} + 2p^{1}_{1,0} - 2p^{1}_{4,0} \leqslant 0$ & 0.5  & 0.2 & 0.8284 & 0 \\
         \hline
         4 & $-2p^{0}_{2,0} -2 p^{0}_{0,1} -2 p^{0}_{1,1} +2p^{0}_{4,1} + 2p^{1}_{2,0} - 2p^{1}_{4,0}\leqslant 0 $ & 0.5  & 0.2 & 0.8284 & 0 \\
         \hline
         2 & $ -p^{0}_{0,0} - p^{0}_{1,0} + p^{0}_{4,0} - p^{0}_{0,1} - p^{0}_{1,1} + p^{0}_{2,1} + p^{1}_{0,0} + p^{1}_{1,0} - p^{1}_{2,0} - p^{1}_{4,0}\leqslant 0 $ & 0.25  & 0.2 & 0.4142 & 0 \\
         \hline
         2 & $ -2p^{0}_{4,0} - 2p^{0}_{2,1} -2 p^{1}_{0,0} - 2p^{1}_{1,0} + 2p^{1}_{2,0} + 2p^{1}_{4,0}\leqslant 0 $ & 0.5  & 0.2 & 0.8284 & 0 \\
         \hline
    \end{tabular}
    \caption{Here, we opt for the notation $p_{x,y}^z = p(z|x,y)$. We obtained $107$ inequalities, out of which $17$ are trivial. The remaining inequalities are reduced to $12$ inequivalent classes that are listed above. In order to identify the equivalent noncontextuality inequalities, the following indistinguishability relations and symmetry transformations are implemented.
    \label{sc9table}}
\small{
$ 
 \ p(z|x=3,y) =  p(z|x=0,y) + p(z|x=1,y) - p(z|x=2,y) \ \forall y,z $
$ 
 \ p(z|x=5,y) =  p(z|x=0,y) + p(z|x=1,y) - p(z|x=4,y) \ \forall y,z 
$ \\
$
\ p(2|x,y) = 1-p(0|x,y) - p(1|x,y) \ \forall x,y
$\\
$
    P_0 \longleftrightarrow P_1 ; 
    P_2 \longleftrightarrow P_3 ;
    P_4 \longleftrightarrow P_5 ;
$\\
$
    P_0 \longleftrightarrow P_2 , 
    P_1 \longleftrightarrow P_3 ;\ 
    P_0 \longleftrightarrow P_4 , 
    P_1 \longleftrightarrow P_5 ; \ 
    P_2 \longleftrightarrow P_4 , 
    P_3 \longleftrightarrow P_5  ; 
$\\
$
    P_0 \longleftrightarrow P_3 , 
    P_1 \longleftrightarrow P_2; \ 
    P_0 \longleftrightarrow P_5 , 
    P_1 \longleftrightarrow P_4 ;\ 
    P_2 \longleftrightarrow P_5 , 
    P_3 \longleftrightarrow P_4 ;
$\\
$  
    \ M_{0|y} \longleftrightarrow M_{1|y}
   \ \forall y ;
    \ M_{z|0} \longleftrightarrow M_{z|1}
   \ \forall z .
$
}
\end{table}
For most of the noncontextuality inequalities, where $\Q^\Pi_1 < \Q^2_s$ and the noncontextual bounds are the same as $\Q^\Pi_1$, any quantum violation certifies nonprojective measurements.
\subsection*{Witnessing quantum violation via qubit strategies}
\begin{figure}[h]
    \centering
    \begin{subfigure}[b]{0.3\textwidth}
        \centering
        \includegraphics[width=\textwidth]{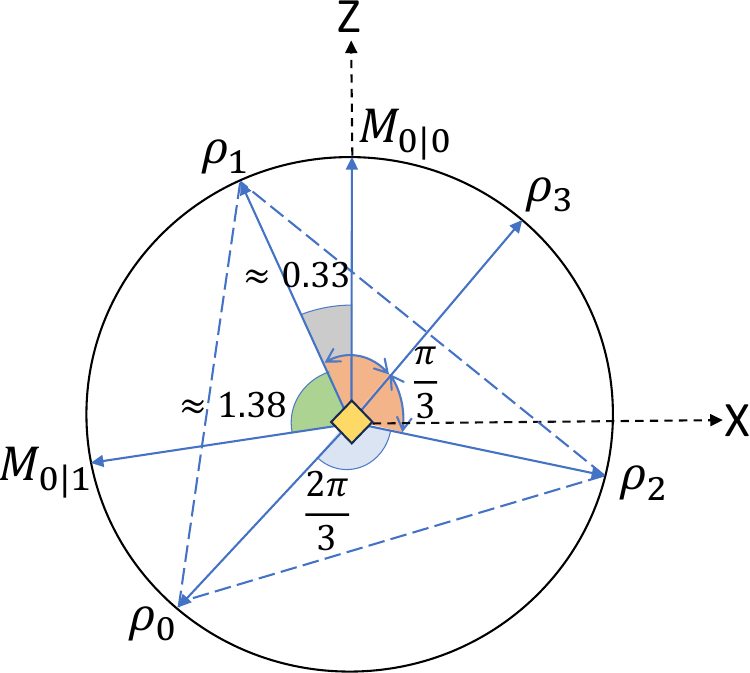}
        \caption{$\mathcal{I}_2 = 2.6458$ (from Tab. \ref{sc2table})}
    \end{subfigure}
    \hfill
    \begin{subfigure}[b]{0.29\textwidth}
        \centering
        \includegraphics[width=\textwidth]{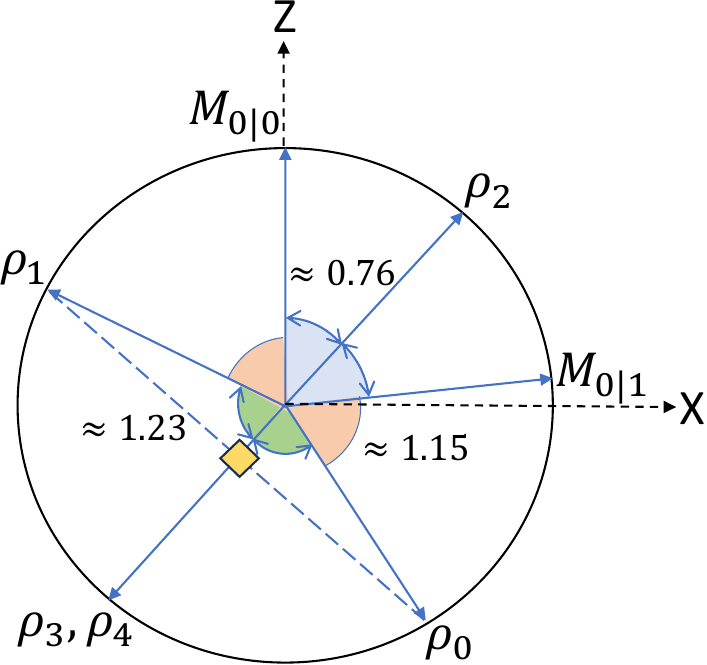}
        \caption{$\mathcal{I}_3 = 3.1231$ (from Tab. \ref{sc3table})}
    \end{subfigure}
    \hfill
    \begin{subfigure}[b]{0.3\textwidth}
        \centering
        \includegraphics[width=\textwidth]{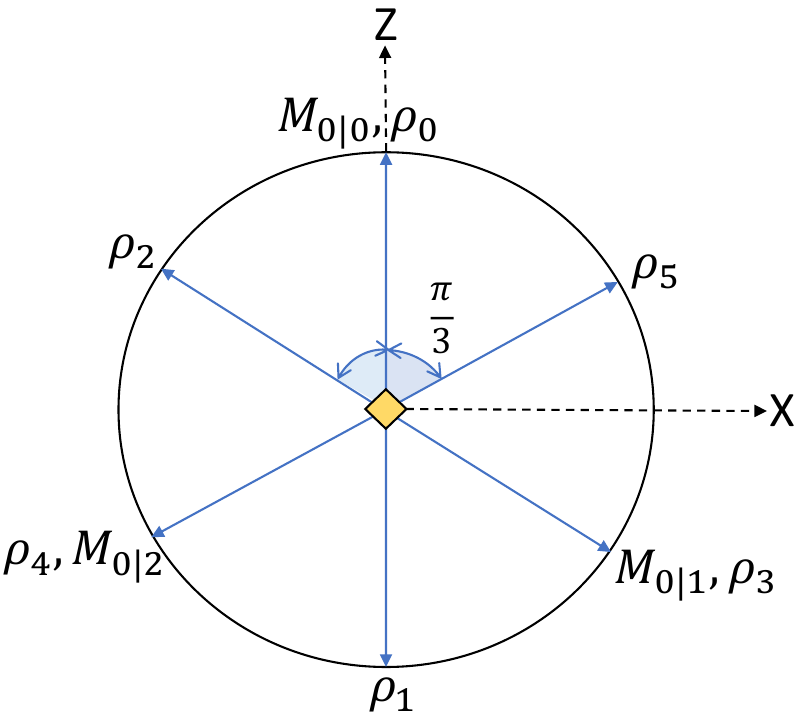}
        \caption{$\mathcal{I}_5 = 2.5$ (from Tab. \ref{sc5table})}
    \end{subfigure} 
    \\
    \begin{subfigure}[b]{0.3\textwidth}
        \centering
        \includegraphics[scale=0.4]{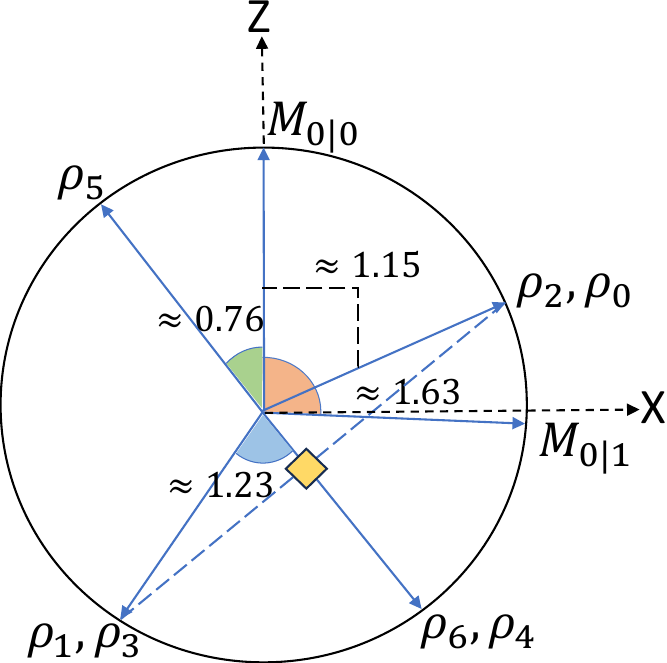}
        \caption{$\mathcal{I}_6^3 = 8.2462$ (from Tab. \ref{sc6table})}
    \end{subfigure}
    \hfill
    \begin{subfigure}[b]{0.3\textwidth}
        \centering
        \includegraphics[scale=0.4]{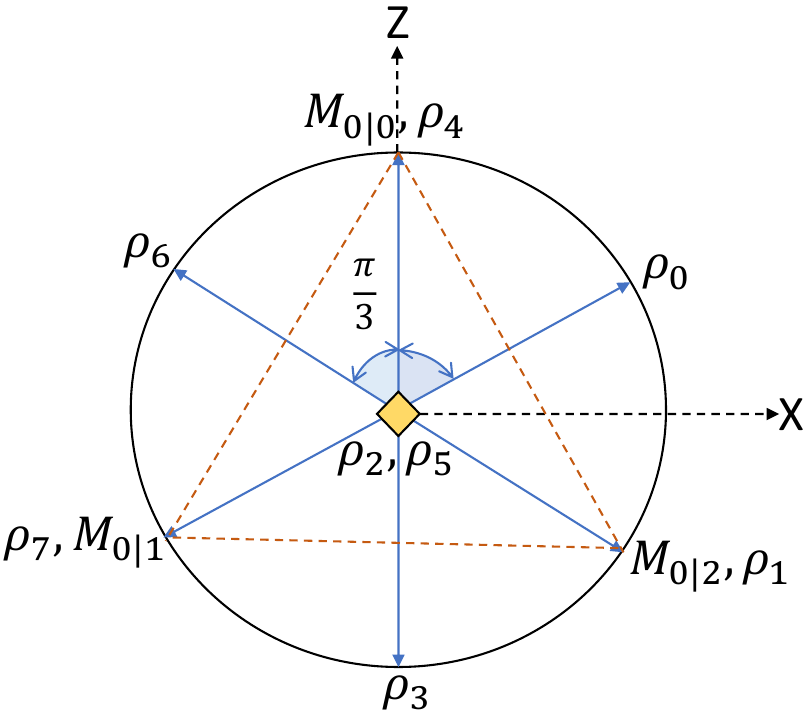}
        \caption{$\mathcal{I}_8 = 1.5$ (from Tab. \ref{sc8table})}
    \end{subfigure}
    \hfill
    \begin{subfigure}[b]{0.3\textwidth}
        \centering
        \includegraphics[scale=0.4]{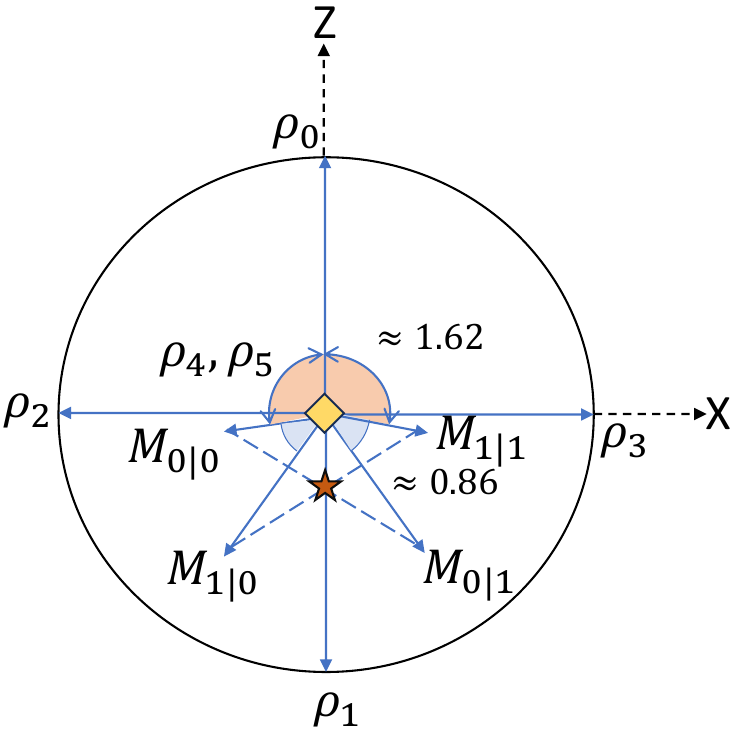}
        \caption{$\mathcal{I}_9 = 1.4536$ (from Tab. \ref{sc9table})}
        \label{sc9fig}
    \end{subfigure}
     \caption{The x-z plane of the Bloch sphere is considered to pinpoint the quantum states and measurements that yield the maximum violations of some of the noncontextuality inequalities, as determined by the see-saw optimization technique for two-dimensional quantum systems. The symbols $\diamond$ and $\star$  represent the indistinguishable mixed state and the indistinguishable measurement effects in the respective scenario. In figure (\ref{sc9fig}), the length of the Bloch vectors representing $M_{0|0}$ and $M_{1|1}$ is approximately $ 0.3689$ and the length of the Bloch vectors representing $M_{1|0},M_{0|1}$ is approximately $ 0.674$.}
    \label{fig:qv}
\end{figure}

Figure \ref{fig:qv} illustrates the qubit strategies attaining the maximum quantum violations. 
In many instances, up to machine precision, the maximal quantum violations have been found whenever $\mathcal{Q}_s^d$ matches with $\mathcal{Q}_1$.
Of all the NCI, the most resilient one emerges from Scenario 7, with the critical robustness parameter equal to $0.423$. 
Intriguingly, the best obtained quantum violations do not always stem from cases where the indistinguishable state is the maximally mixed state. Scenario $6$, which features seven preparations and binary outcomes, represents the simplest scenario showcasing variations in quantum violations for different dimensional quantum systems. 

Based on the see-saw optimization, a range of noncontextuality inequalities exist, violations of which serve as witnesses of the dimension of the quantum systems. For instance, if a quantum violation of $\mathcal{I}_6^1$ surpasses $7.6833$, then the dimension of the quantum systems must be at least three. Moreover, Scenario $9$ provides several noncontextuality inequalities whose violations using qubits certify three outcome non-projective measurements. In fact, non-projective measurements are necessary for achieving quantum violations of all the noncontextuality inequalities in this contextuality scenario.

\twocolumngrid

\section{Applications of newly found noncontextuality inequalities}

In this section, we delve into some of the interesting applications of our newly  discovered noncontextuality inequalities.

\subsubsection*{Quantum advantage in oblivious communication}

Oblivious transfer is a crucial task in information theory with myriad applications in cryptography. 
In \cite{SahaNJP}, the oblivious transfer task has been generalized, and it is shown that any quantum advantage in such tasks, namely, oblivious communication task, implies preparation contextuality. Moreover, quantum violations of preparation noncontextuality inequalities in prepare-and-measure scenarios directly translate into quantum advantages in oblivious communication tasks. Here, the oblivious condition is associated with the indistinguishable condition on the preparations, and the expression of the respective noncontextuality inequality corresponds to the figure of merit of that task. Consequently, any quantum violation of newly discovered preparation noncontextuality inequalities can be interpreted as a quantum advantage in an oblivious communication task.

In light of our present analysis, it may be worthwhile to mention the
following examples.
noncontextuality inequality obtained from the simplest contextuality scenario discussed in Section III serves as  the success metric of the parity oblivious random access codes \cite{Spekkens2009}. Next, an interesting fact emerges from the discussion after Table \ref{sc2table}. The optimal classical encoding strategy for the sender in the oblivious communication task with respect to $\mathcal{I}_2$ must be probabilistic for saturating the noncontextual bound $2$, and the maximum value of $\mathcal{I}_2$ for deterministic encoding strategies is $1$. Further, as already noted earlier, the inequality $\mathcal{I}_7$ 
can be employed for the 3-bit parity oblivious multiplexing task \cite{Spekkens2009}. Moreover, the optimal classical encoding strategy for the sender in the oblivious communication task with respect to $\mathcal{I}_6^2$ is bounded by $12$, and a quantum advantage ensues whenever  $\mathcal{I}_6^2 > 12$.

\subsubsection*{Certification of non-projective measurements}

The study by Chaturvedi \textit{et al.} \cite{chaturvedi2021quantum} points out that the maximum value of a noncontextuality inequality can always be achieved using projective measurement, where no indistinguishability conditions on measurements are imposed. Consequently, certification of non-projective measurements through the violation of noncontextuality inequalities can only occur with nontrivial indistinguishable conditions on measurements. To accomplish this, it is necessary to establish an upper bound on the expression of noncontextuality inequalities when measurements are restricted to be projective for arbitrary dimensional quantum states and measurements. As outlined in \cite{Chaturvedi2021characterising}, the semi-definite hierarchy can be modified to obtain upper bounds when the measurements are projective. Let us denote these upper bounds by $\Q_1^{\Pi}$. 

We have implemented this optimization to obtain $\Q_1^{\Pi}$ in the last two scenarios under consideration, both featuring nontrivial indistinguishability conditions for measurements. Among these scenarios, it was found that in scenario VIII, $\mathcal{Q}_1$ is the same $\Q_1^{\Pi}$ for all noncontextuality inequalities, suggesting this scenario is unable to certify non-projective measurements. However, in scenario IX, the $\Q_1^{\Pi}$ values are lower than $\mathcal{Q}_s^2$ for all noncontextuality inequalities whenever a quantum violation occurs. These precise values are documented in the final column of Table \ref{sc9table}. For example,
corresponding to quantum violation of $\mathcal{I}_9$,  the lower bound $\Q^2_s = 1.453$ and the upper bound 
$\Q_1 =  2.828$  are obtained. Further, $\Q_1^{\Pi} = 1$, certifying 
non-projectiveness of the measurements. 
Notably, the upper bounds for projective measurement coincide with the noncontextual values for all noncontextuality inequalities, indicating that any violation of these noncontextuality inequalities implies unequivocally that the measurements are non-projective.


\subsubsection*{Noncontextuality inequalities as dimension witnesses}

As a consequence of our extensive investigations, we reveal a noteworthy aspect of the interplay between quantum preparation contextuality and quantum Hilbert space dimension.  Specifically, we implement a \emph{novel} hierarchy of SDP relaxations. In particular, the novel scheme employs the scheme described in \cite{Chaturvedi2021characterising} over the 
basis generated from the Navascu\'es-V\'ertesi method for bounding finite-dimensional quantum correlations and retrieve tight upper bounds $Q^2_{UB}=Q^{2}_{s}$ with level $3$ for inequalities in Table \ref{sc6table}. 
Moreover, without the operational equivalences, the dimension restriction fails to yield non-trivial bounds on the inequalities; the operational equivalences are \emph{necessary} to witness the Hilbert dimension $d>2$ with these noncontextuality inequalities.  

Notably, the lower bounds obtained from the see-saw for dimension $d>2$ violate the upper bounds for seven noncontextuality inequalities, thereby forming hitherto unknown non-trivial dimension witnesses.
In particular, we find that \emph{seven} of ten noncontextuality inequalities in Scenario 6, Table \ref{sc6table}, double as dimension witnesses for Hilbert space dimension $d>2$. It is worth mentioning that $\mathcal{I}_6^2$ in Tab.\ref{sc6table} is violated by qutrit systems, while qubit states fail to produce any violation. 

\subsubsection*{Randomness certification}
Device-independent (DI) randomness certification from Bell-inequality violation requires spacelike separation and is experimentally demanding. Noncontextuality-based randomness certification offers a \emph{semi–device-independent middle ground}: devices remain uncharacterised except for verifiable preparation equivalences. Conceptually, it probes a different boundary--- that of \emph{quantum contextual behavior under specified operational equivalences}---and can certify randomness in \emph{local} prepare-and-measure scenarios where DI certification is impossible. Thus the appeal is both \emph{operational} (easier implementations) and \emph{theoretical} (a distinct notion of classicality being excluded and fueling the certification of randomness). 

Here we demonstrate that novel instances of quantum contextuality found 
through our approach can be used for semi-device-independent quantum randomness certification with operational equivalences. To exemplify this we consider a prepare-and-measure setting with a sender (Alice) and a receiver (Bob). Upon receiving a uniformly random classical input $x\in\{0,1,2,3,4,5,6,7\}$, Alice’s device emits a quantum state $\rho_x\in B_+(\mathcal H)$. Upon receiving a uniformly random input $y\in\{0,1,2\}$, Bob’s device performs a binary POVM $\{M_{0|y},\,M_{1|y} \in B_+(\mathcal H)\}$ with $M_{1|y}=\mathbb I_{\mathcal H}-M_{0|y}$. We adopt an \emph{extreme adversarial} (untrusted) model in which an eavesdropper Eve may have designed and have complete control over both devices, i.e., she can choose and modify $\{\rho_x\}$, $\{M_{0|y}\}$, and the underlying Hilbert space $\mathcal H$. The \emph{sole} assumption is that Alice’s preparations satisfy the operational equivalences \eqref{sc7a}. This constitutes a \emph{semi–device-independent} assumption.

Instead of using multi-parameter observed statistics, the parties consider a single parameter, namely the value of the non-contextuality inequality $\mathcal{I}_7$ in Table \ref{sc7table} associated with Scenario 7. Suppose the experiment yields a violation $\mathcal I_7=i_7>1$. Let Alice and Bob extract randomness from the fixed setting $(x^\star,y^\star)=(0,0)$. Eve’s optimal guessing probability is then the solution of
\begin{align} 
p^{\ast}_{\mathrm{guess}}(i_7)
:=\ \max_{\{\rho_x\},\,\{M_{0|y}\},\,\mathcal H}\ \ & 
\operatorname{Tr}\!\left[\rho_{x^\star}M_{0|y^\star}\right]\nonumber\\[2pt]
\text{s.t.}\quad &
0\preceq \rho_x,\ \forall x;\nonumber  \\
&
0\preceq M_{0|y}\preceq \mathbb I_{\mathcal H},\ \forall y;\nonumber\\
& \{\rho_x\} \text{ satisfy \eqref{sc7a} };\nonumber\\
& \mathcal I_7(\{\rho_x\},\{M_{0|y}\})\,=\,i_7.
\label{eq:opt-prob}
\end{align}
The certified per-round min-entropy is
\begin{equation} 
h(i_7)\;=\;-\log_2 p^{\ast}_{\mathrm{guess}}(i_7).
\label{eq:min-entropy}
\end{equation}

Because $\mathcal H$ is unbounded and constraints are noncommutative, solving \eqref{eq:opt-prob} exactly is hard. We therefore use the noncommutative polynomial-optimisation hierarchy of Ref.~\cite{Chaturvedi2021characterising} to obtain an \emph{upper bound} $p^{*'}_{\mathrm{guess}}(i_7)\ge p^{\ast}_{\mathrm{guess}}(i_7)$, yielding a \emph{lower bound} $h'(i_7)=-\log_2 p^{*'}_{\mathrm{guess}}(i_7)$ on certifiable randomness.

\begin{figure}[!ht]
    \centering
    \includegraphics[scale=0.43]{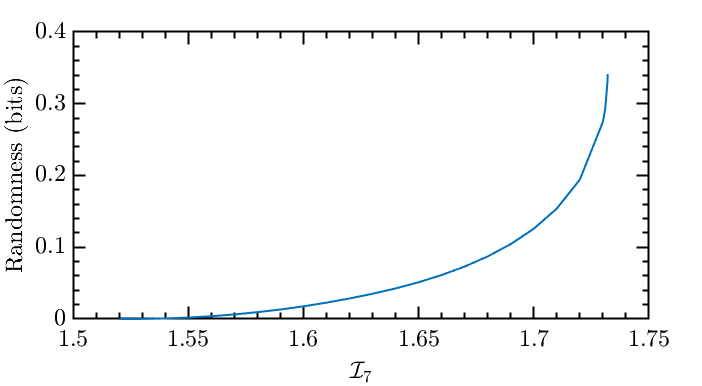}
    \caption{Randomness ($H_{min}$) as a function of $\mathcal{I}_7 \in [1.5,1.7321]$ from Table \ref{sc7table}.}
    \label{Randomness3To1}
\end{figure}

In FIG. \ref{Randomness3To1} we plot the lower bounds on certified randomness $h'(i_7)$ obtained from level $3$ of the SDP hierarchy against the violation of $\mathcal{I}_7$. We find that non-zero randomness can be certified in the range $\mathcal{I}_7\in [1.52,1.7321]$ with maximum $0.34147$ coinciding with the maximum attainable violation $\mathcal{I}_7= Q^2_s = Q_1$ (up to machine precision). We note here that the choice of $x=0,y=0$ turns out be optimal and equivalent to $x=0,y=1$ and $x=0,y=2$. 

\section{Conclusions} 

Deriving a set of empirical criteria applicable to any operational theory that satisfies the generalized notion of contextuality is an arduous task of both foundational and operational significance. The conventional method \cite{Schmid18} of extracting facet inequalities from the pertinent noncontextual polytope is computationally demanding due to the exponential growth in the dimension of the polytope describing the preparations with the number of measurements. In this work, we introduce a computationally efficient procedure for constructing the noncontextual polytope. An approach employing flag-convexification to characterize the noncontextual polytope in general scenarios involving preparations, transformations and measurements has recently been proposed by Schmid \textit{et al.} \cite{schmid2024noncontextualityinequalitiespreparetransformmeasurescenarios}. While the essential elements of our approach were formulated prior to the
appearance of Ref. \cite{schmid2024noncontextualityinequalitiespreparetransformmeasurescenarios},    connections and comparisons between these two nearly concurrently proposed approaches  remain as interesting questions for future study. 

We demonstrate the efficacy of our proposed method by applying it to 
several examples of contextuality scenarios.  Consequently, we retrieve a large number of novel noncontextuality inequalities, violations of which serve as necessary and sufficient conditions for demonstrating quantum contextuality in these scenarios. We employ two semi-definite programming techniques for retrieving
the lower and upper bounds, respectively, on the maximum quantum violations
of the noncontextuality inequalities.  We further study the robustness of the quantum violations against noise. Among the obtained noncontextuality inequalities, $\mathcal{I}_7$ turns out to be the most robust inequality. Our investigation uncovers hitherto unexplored non-trivial noncontextuality inequalities and reveals intriguing aspects of quantum contextual correlations, including applications in information processing tasks such as oblivious communication, dimension witness, certification of non-projective measurements and randomness generation. Another interesting observation concerns the certification of epistemic randomness using noncontextuality inequalities (for example, the case of $\mathcal{I}_2$). If epistemic randomness is restricted, then the achievable value of such noncontextuality inequalities can be lower than the actual noncontextual bound. This feature is distinct from other notions of classicality and opens up new possibilities to test a hierarchy between determinism, epistemic randomness, and quantum randomness in contextual scenarios.

The present study has focused on sets of indistinguishability conditions regarding preparations and measurements, respectively, to render them indistinguishable from each other. It is possible to consider scenarios with more than one set of indistinguishability conditions for a given scenario, each corresponding to convex decompositions of mixed preparations or measurements. Extending our method to cover such scenarios could be explored more thoroughly in future research. In the future, it would be interesting to extend this study to general probabilistic theories to examine how the noncontextuality inequalities derived in this work are violated beyond the quantum framework. The inherently contextual nature of quantum theory offers several distinct advantages in cryptographic and computational tasks. 
Our present analysis 
should motivate future endeavours to leverage newfound instances of quantum contextuality for a wide range of information-theoretic applications.

\subsection*{Acknowledgment} We thank the anonymous referees for their helpful suggestions. DS acknowledges financial support from STARS (STARS/STARS-2/2023-0809), Govt. of India. AC acknowledges financial support by NCN Grant SONATINA 6 (Contract No. UMO-2022/44/C/ST2/00081). SB acknowledges financial support from the National Science and Technology Council, Taiwan (Grants No. 12-2628-M-006-007-MY4, 115-2811-M-006-002-MY2).

\bibliographystyle{alphaarxiv}

\bibliography{ref}

\end{document}